\documentclass[%
 aip,
 amsmath,amssymb,
preprint,%
]{revtex4-1}
\usepackage{mathrsfs,amsmath,amssymb,bm}
\usepackage{hyperref}
\usepackage{graphicx}
\usepackage{subfigure}
\usepackage{caption}
\usepackage{overpic}
\usepackage{epsfig}
\usepackage{xcolor}
\usepackage{indentfirst}
\usepackage{fancyhdr}
\usepackage{comment}
\usepackage{makecell, rotating}
\usepackage{algorithm} 
\usepackage{algorithmic} 
\usepackage{amsthm}

\theoremstyle{remark}

\begin{document}


\title{Analytic Structure of the SCFT Energy Functional of Multicomponent Block Copolymers}

\author{Kai Jiang}
 \email{kaijiang@xtu.edu.cn}
 \affiliation{Hunan Key Laboratory for Computation and Simulation in Science
and Engineering, School of Mathematics and Computational Science,
Xiangtan University, Hunan, P.R. China, 411105}
 \affiliation{LMAM, CAPT and School of Mathematical Sciences, Peking University, Beijing 100871, P.R. China}

\author{Weiquan Xu}
\affiliation{LMAM, CAPT and School of Mathematical Sciences, Peking University, Beijing 100871, P.R. China}

 \author{Pingwen Zhang}
\thanks{}
 \email{pzhang@pku.edu.cn}
 \affiliation{LMAM, CAPT and School of Mathematical Sciences, Peking University, Beijing 100871, P.R. China}
\homepage{http://www.math.pku.edu.cn/pzhang}

\date{\today}

\begin{abstract}
	This paper concerns the analytic structure of the
	self-consistent field theory (SCFT) energy functional of
	multicomponent block copolymer systems which contain more
	than two chemically distinct blocks. The SCFT has enjoyed
	considered success and wide usage in investigation of the
	complex phase behavior of block copolymers. 
	It is well-known that the physical solutions of
	the SCFT equations are saddle points, however, the analytic
	structure of the SCFT energy functional has received little
	attention over the years.
	A recent work by Fredrickson and collaborators [see the
	monograph by Fredrickson, \textit{The Equilibrium Theory of
	Inhomogeneous Polymers}, (2006), pp. 203--209] 
	has analysed the mathematical structure of the field
	energy functional for polymeric systems, and
	clarified the index-1 saddle point nature of the problem produced by
	the incompressibility constraint. In this paper, our goals
	are to draw further attention to multicomponent block
	copolymers utilizing the Hubbard-Stratonovich transformation
	used by Fredrickson and co-workers. We first
	show that the saddle point character of the SCFT energy
	functional of multicomponent block copolymer systems may be
	high index, not only produced by the
	incompressibility constraint, but also by the Flory-Huggins
	interaction parameters. Our analysis will be beneficial to
	many theoretical studies, such as the nucleation theory of
	ordered phases, the mesoscopic dynamics. As an application,
	we then utilize the discovery to develop the gradient-based
	iterative schemes to solve the SCFT equations, and illustrate
	its performance through several numerical experiments taking
	$ABC$ star triblock copolymers as an example.
\end{abstract}

\maketitle


\section{Introduction}

Due to the efforts of a large number of researchers, it has
been well-established that the self-consistent field theory
(SCFT) of polymers provides a powerful theoretical framework for
the study of inhomogeneous polymeric systems in general and the
self-assembly behavior of block copolymers in
particular\,\cite{hamley2004developments, matsen2002standard,
fredrickson2006equilibrium}. For a given block copolymer system, the
SCFT can be efficiently describe its architecture,
molecular composition, polydispersity, and block types as a
series of parameters in the energy functional which is a 
nonlinear and nonlocal functional of the monomer densities and
their conjugate fields. The equilibrium solutions of the SCFT
energy functional correspond to the possible stable and
metastable phases of the block copolymer system. They
are determined by a set of the SCFT equations obtained by
assigning the first variations of energy functional with respect
to the density profiles and fields to zero. 
Finding all solutions of the SCFT equations analytically is
beyond today's technology, even for the simplest $AB$ diblock
copolymer system. A successful alternative is to solve
the SCFT equations numerically.

Numerically solving the SCFT equations requires the analysis of 
the property of SCFT solutions, and even the mathematical
structure of the SCFT energy functional. 
There have been a number of numerical techniques developed to solve
SCFT equations, including from the perspective of three facets --
the strategy of screening initial
values\,\cite{jiang2010spectral, jiang2013discovery,
xu2013strategy}, the numerical solver for the modified diffusion
equation of chain propagators\,\cite{vavasour1992self, matsen1994stable,
drolet1999combinatorial, rasmussen2002improved, guo2008discovering,
cochran2006stability, ranjan2008linear}, and the iterative
schemes for the convergence of the equations
system\,\cite{drolet1999combinatorial, thompson2004improved,
ceniceros2004numerical, liang2013efficient}. 
However, little work has been devoted to further analysing the
mathematical structure of the SCFT even through it is well-known
that the equilibrium solutions of the SCFT equations are
saddle points\,\cite{fredrickson2006equilibrium}. Among these
researches, Fredrickson and co-workers\,\cite{ganesan2001field,
fredrickson2002field, fredrickson2006equilibrium} have directly
analysed the analytic structure of the field-based energy
functional which does not use the mean-field approximation for
incompressible polymeric systems.
They utilized the Hubbard-Stratonovich
transformation\,\cite{chaikin1995principles} to decouple the
particle-particle interaction, and pointed out that, for binary
component system, the physical
solutions represent saddle points in which the energy
functional is minimized with respect to the exchange chemical
potential, and maximized with respect to the pressure potential.
It is the latter field that imposes the incompressibility
constraint and produces the index-1 saddle point character of the
problem. 
The analysis of the mathematical structure of the SCFT
energy functional, or more generally, of the field functional,
is very useful for the theoretical studies of polymer systems.
For this reason, a series of theoretical tools have
been developed including the efficient gradient-based iterative
methods to solve the SCFT
equations\,\cite{ceniceros2004numerical, liang2013efficient}, the
useful technique of partial saddle-point approximation
translating the saddle point problem into extreme value
problem\,\cite{reister2001spinodal, duchs2003fluctuation}, and
the solving strategy for field theoretic simulation which
involves numerically solving the exact partition function of
polymer fluid models\,\cite{ganesan2001field, fredrickson2002field}.
Due to the known information of the saddle point character,
the SCFT model can be applied to the mean-field
mesoscopic dynamics\,\cite{fraaije1993dynamic,
fraaije1997dynamic}, and the nucleation theory of ordered
phases in which the saddle point problem should be turned into a
minimum problem firstly, followed by the string
method\,\cite{cheng2010nucleation}.

On the other hand, in recent years, a large number of studies
have been performed on the investigation of the phase behavior of
multicomponent block copolymers because more chemically distinct
blocks can offer opportunities to create a greater diversity of
order phases and phase behaviors\,\cite{bates2012multiblock,
masuda2006nanophase, epps2004ordered, tang2004star,
guo2008discovering, xu2013strategy}. 
Because of the rapidly increasing parameters charactered the
multicomponent block copolymer, the SCFT energy functional
becomes more complex which also leads to a larger scale of SCFT
equations system. This increases the difficulty of theoretical
research and numerical simulations, and requires more theoretical
analysis of the mathematical structure of the SCFT energy functional.
For example, the recently developed nucleation theory of ordered
phases within the SCFT using string
method\,\cite{cheng2010nucleation}, it involves converting the
saddle point problem of SCFT into a minimal value problem in each
iteration by a priori knowledge of saddle point character. This
approach has been restricted to diblock copolymers because of
lack of analysis on the saddle point property of the SCFT energy
functional of multicomponent copolymer systems.
In the present paper, we will analyse the analytic structure of
the SCFT energy functional of multicomponent
block copolymer systems based on the work of Fredrickson
\emph{et al.}. The adopted technique to use the
Hubbard-Stratonovich transformation to decouple the
particle-particle interaction, is not new, but it is the first
time to analyse the SCFT energy functional of multicomponent
(more than $2$) block copolymers. The analytic structure of
multicomponent block copolymer system is different from that of
two-component polymer systems.
In the latter system, the saddle point of physical solution is caused by 
the incompressibility constraint\,\cite{ganesan2001field,
fredrickson2002field, fredrickson2006equilibrium}.
However, for multicomponent polymer systems, except for the
incompressibility constraint, the existence of multiple
non-bonded interactions will lead to high index saddle points
with certain Flory-Huggins interaction parameters. In this work 
the influence of Flory-Huggins interaction parameters
for saddle point character of the SCFT will be detailed analysed. 

The organization of the article is as follows. 
In Sec.\,\ref{sec:theory}, we present the theoretical framework
of the SCFT for multicomponent block copolymer system. Meanwhile 
the saddle point properties of the SCFT energy functional will be
analysed detailedly, especially considering the influence of
different Flory-Huggins interaction parameters. We also take the
$ABC$ triblock copolymers as a specific example to
illustrate our theory in this section.
In Sec.\,\ref{sec:method}, as an application, we will extend the
gradient-based iterative method, developed in diblock copolymers,
to solve the SCFT equations of multicomponent block copolymer system.
Sec.\,\ref{sec:results} shows the corresponding numerical results to
demonstrate the availability of our developed model and numerical
methods, using $ABC$ star triblock copolymers model.
The final section presents brief summary and some discussions on the
proposed theory.

\section{Theory}
\label{sec:theory}

\subsection{Statistical particle theory}
\label{subsec:statisticalParticle}

The first ingredient in the SCFT of a polymeric fluid is a
mesoscopic molecular model to describe the statistical mechanics
associated with the conformational states of a single polymer.
In order to describe the properties of polymer chains, lots of
coarse-grained polymer chain models have been developed for
theoretic studies, such as continuous Gaussian
chain model for flexible polymers, wormlike chain model for
semiflexible or rigid-rod polymers\,\cite{rubinstein2003polymer}.
For illustrative purpose, in this paper, we consider an
incompressible canonical ensemble system with volume $V$ of $n$
flexible block copolymers.
Each block copolymer has $K$ chemically distinct
subchains and $M$ blocks (For example, $ABA$ triblock
copolymer has two chemically distinct subchains, $K=2$, but
three blocks, $M=3$).
The continuous Gaussian
chain model is utilized to describe macromolecular conformations.
The total degree of polymerization of a block copolymer is $N$,
and the degree of polymerization of the $m$-th block
is $N_m$, then $N = \sum_{m=1}^{M} N_m$.
In the continuous Gaussian model, block copolymers are
represented by continuous space curves $\bm \Gamma^j(s), j = 1,
2, \dots, n$, where $s \in [0,1]$ is a normalized arc length variable measured
along the chain contour. These curves are not necessary to be 
linear, and are also allowed the existence of star, graft shaped, and
even more complex topological structure. 
For the $j$-th polymer chain, $\bm
\Gamma^j_\alpha(s)$ ($s\in I_{\alpha}$) denotes the blocks whose
monomers belong to the same species $\alpha, \alpha \in \left\{ A, B, C,
\dots \right\}$, and $I_{\alpha}$ is the interval of
corresponding contour parameter. $\bm \Gamma^j_\alpha(s)$ ($s\in
I_{\alpha}$) may be discontinuous. For example, for $ABA$ linear
triblock copolymer, $\bm \Gamma^j_A(s)$ is used to describe the
configuration of the left block $A$ and the right block $A$ of
the $j$-th polymer chain.

The normalized microscopic monomer density of the type-$\alpha$
species at space position $\bm r$ is
\begin{align}
	\hat \rho_{\alpha}(\bm r) = \frac{N}{\rho_0} \sum_j \int_{I_{\alpha}} \mathrm
ds \delta\left( \bm r - \bm \Gamma^j_\alpha(s) \right).
	\label{}
\end{align}
The average monomer density is $\rho_0 = nN/V$.
Without geometry confinement or external fields, the energy
functional
$H[\bm \Gamma]$, $\bm \Gamma = \{\bm \Gamma^1, \bm\Gamma^2, \dots,
\bm\Gamma^n\}$, of a specific configuration state of the
copolymer melt system contains harmonic stretching energy
$U_0[\bm \Gamma]$ and non-bonded interactions of between distinct
monomer species $U_1[\bm
\Gamma]$\,\cite{fredrickson2006equilibrium}
\begin{equation}
  \begin{split}
    H[\bm \Gamma] &= U_0[\bm \Gamma] + U_1[\bm \Gamma],\\
    \beta U_0[\bm \Gamma] &= \frac{3}{2N b^2}\sum_{j=1}^{n}
	\int_0^1 \mathrm{d} s \Big|\frac{\mathrm{d} \bm
	\Gamma^j(s)}{\mathrm{d} s}\Big|^2,\\
    \beta U_1[\bm \Gamma] &= \rho_0 \int \mathrm{d} \bm
	r~\sum_{\alpha \neq \beta} \chi_{\alpha \beta} \hat
	\rho_{\alpha}(\bm r) \hat \rho_{\beta}(\bm r),
  \end{split}
  \label{eq:coarsegrainedenergy}
\end{equation}
where $\chi_{\alpha \beta}$ is the Flory-Huggins interaction
parameters. We here assume all statistical segments occupy the
same volume $v_0$, and ignore differences in statistical segment
length among all blocks, i.e., $b_\alpha=b$.
Incompressibility of the system is ensured by
the delta functional in the partition function
\begin{equation}
  Z = \int \mathcal D [\bm \Gamma]~\delta [\hat \rho_+(\bm r) - 1] \times \exp \left( -\beta H[\bm \Gamma] \right),
  \label{eq:CoarseGrainedPartitionFunctional}
\end{equation}
where $\hat \rho_+(\bm r) = \sum_{\alpha} \hat\rho_{\alpha}(\bm
r)$ denotes the total segment density.  Thermal equilibrium
occurs when the distribution function of microscopic
configuration of block copolymer melts can stabilize to the
Maxwell-Boltzmann distribution.  In other words, the probability
of finding a specific microscopic configuration $\bm \Gamma$ is
\begin{equation}
  P[\bm \Gamma] = \frac{\exp\left( -\beta H[\bm \Gamma] \right)}{Z},
  \label{eqn:MBdist}
\end{equation}
with the configuration $\bm \Gamma$ subjected to the
incompressible condition $\hat \rho_{+}(\bm r) = 1$.

The essence of the saddle point (mean-field) approximation is to
evaluate the most possibly observed configuration, or
equivalently dominant configuration corresponding to the global
minima of the energy functional $H[\bm \Gamma]$ subjected to the
constraint of incompressibility, and neglect all the
fluctuations. For the convenience of subsequent studies of
evaluating the dominant configuration and approximated
energy functional, we are going to transform this particle-based 
statistical mechanism into the statistical field theory mainly
through the Hubbard-Stratonovich transformation (also named as 
Gaussian functional integration technique) that serves to
decouple particle-particle
interactions\,\cite{chaikin1995principles,
fredrickson2006equilibrium}.

\subsection{Statistical field theory}
\label{subsec:statisticalField}

In order to use the Hubbard-Stratonovich transformation, 
we have to first transform the non-bonded interaction $U_1[\bm
\Gamma]$ into the normalized form
\begin{equation}
  \beta U_1[\bm \Gamma] = - \rho_0 \int \mathrm d \bm r~ \sum_{k} \zeta_k \cdot \hat \rho_k^2,
  \label{eq:NormInteraction}
\end{equation}
$\zeta_k$ is the combination of the Flory-Huggins interaction
parameters $\chi_{\alpha\beta}$
and $\hat \rho_k = \sum_{\alpha} \sigma_{k \alpha} \hat
\rho_{\alpha}$ is the linear composition of microscopic monomer
densities $\hat \rho_{\alpha}, \alpha \in \{A, B, C,
\dots\}$. Applying local incompressibility condition $\hat \rho_+
= 1$, or equivalently $\hat \rho_A(\bm r) = 1 - \sum_{\alpha
\neq A} \hat \rho_{\alpha}$, to the integrand of $U_1[\bm
\Gamma]$ in \eqref{eq:coarsegrainedenergy}, we have
\begin{equation}
  \sum_{\alpha \neq \beta} \chi_{\alpha \beta} \hat \rho_{\alpha}(\bm r) \hat \rho_{\beta}(\bm r) = G\left[ \hat \rho_B(\bm r), \hat \rho_C(\bm r) \dots \right] + L\left[ \hat \rho_B(\bm r), \hat \rho_C(\bm r) \dots \right],
  \label{eq:QuadraticForm}
\end{equation}
where $G$ is a quadratic form and $L$ is a linear composition of
$\{\hat \rho_B, \hat \rho_C, \dots\}$ (Appendix
\ref{app:interEnergy} gives the derivation process).
As is well known,
quadratic form $G$ can be normalized as $\sum_{k} -
\zeta_k \cdot \hat \rho_k^2$. Thus the non-bonded interaction
energy is reduced
to $ \beta U_1[\bm \Gamma] = - \rho_0 \int \mathrm d \bm r~
\sum_{k} \zeta_k \cdot \hat \rho_k^2$ up to a constant.
Because $\int \mathrm d \bm r~\hat \rho_{\alpha}(\bm r)$ is a
constant unrelated to polymer configuration, adding any linear
composition of $\hat \rho_{\alpha}(\bm r)$ to the integrand only
makes a constant shift to the interacting energy $U_1[\bm \Gamma]$.
Furthermore, because of the local incompressibility, $\left\{
\zeta_k, \hat \rho_k \right\}$ and $\left\{ \zeta_k, \hat {\tilde
\rho}_k = \sum_{\alpha} \tilde \sigma_{k \alpha} \hat
\rho_{\alpha} = \hat \rho_k - \frac{1}{K} \sum_{\alpha} \sigma_{k
\alpha} \hat \rho_{+} \right\}$ describe the same interacting
energy $U_1[\bm \Gamma]$ up to a constant, but $\hat {\tilde
\rho}_k$ is orthogonal to $\hat \rho_{+}$ for $\sum_{\alpha}
\tilde \sigma_{k \alpha} = 0$.
From our experience, the formula $\left\{ \zeta_k, \hat {\tilde
\rho}_k \right\}$ has a better numerical performance.
There are plenty of choices for coefficients $\zeta_k$ and linear
composition $\hat \rho_k$, however, these expressions essentially
describing the same non-bonded interaction $U_1[\bm \Gamma]$ with
the difference from each other being a constant. 

We apply the Fourier transformation to delta functional
\eqref{eq:deltafourierLM}
\begin{equation}
  \delta[\rho_+ - 1] = \int \mathcal D[W_+] \exp\left( -i \int
  \mathrm d\bm r W_+ \cdot (\rho_+ - 1) \right),
  \label{eq:deltafourierLM}
\end{equation}
and the Hubbard-Stratonovich
transformation (see the Appendix\,\ref{app:gauss}) to the
``particle-based'' partition function
\eqref{eq:CoarseGrainedPartitionFunctional} and have
\begin{equation}
  \begin{split}
    Z = \int \mathcal D[\bm \Gamma] &\int \mathcal D[W_+] \int \prod_{k} \mathcal D[W_k] \exp\Bigg\{ - \int \mathrm d \bm r \sum_{\zeta_k > 0} \left( \frac{1}{4\zeta_k \rho_0} \cdot W_k^2 - W_k \cdot \hat \rho_k \right)\\
	&- \int \mathrm d \bm r \sum_{\zeta_k < 0} \left( \frac{1}{4\zeta_k \rho_0} \cdot W_k^2 - i W_k \cdot \hat \rho_k \right)
	+ \int \mathrm d \bm r iW_+(\hat \rho_+ - 1) - \beta U_0[\bm \Gamma] \Bigg\}.
  \end{split}
  \label{eq:CoarseGrainedPartitionFunctional2}
\end{equation}
Integrating out the chain configuration $\int \mathcal D[\bm
\Gamma]$ leads the particle-based partition function into
field-based formulism
\begin{align}
    Z &= \int \mathcal D[W_+] \int \prod_{k} \mathcal
	D[W_k] \exp\left( -\beta H[\mu_+, \bm \mu] \right), ~~~
	\bm \mu = (\mu_1, \mu_2, \dots),
  \label{eq:FieldBasedPartitionFunctional2}
	\\
    H[\mu_+, \bm \mu] &= \frac{n}{V} \int \mathrm d \bm r \left[ - \mu_+
	+ \sum_{\zeta_k > 0}\frac{1}{4\zeta_kN}\mu_k^2 -
	\sum_{\zeta_k < 0}\frac{1}{4\zeta_kN}\mu_k^2 \right] - n \log
	Q[\bm \omega], ~~~ \bm\omega = (\omega_A, \omega_B, \dots),
  \label{eq:Hamiltonian2}
	\\
    \mu_+ &= \frac{iN}{\rho_0} W_+, \hspace{2mm} \mu_k = \frac{iN}{\rho_0} W_k \hspace{2mm} (\text{if} \hspace{2mm} \zeta_k < 0), \hspace{2mm} \mu_k = \frac{N}{\rho_0} W_k \hspace{2mm} (\text{if} \hspace{2mm} \zeta_k > 0),
	\\
    \omega_{\alpha} &= \mu_+ - \sum_{k} \sigma_{k\alpha} \mu_k,
	\hspace{2mm} \alpha \in \left\{ A, B, C, \dots \right\},
  \label{eq:fieldsRelation}
\end{align}
where $Q[\bm \omega]$ is the partition function of a single copolymer
under effective mean field $\omega_{\alpha}$ on type-$\alpha$ species,
$\alpha \in \left\{ A, B, C, \dots \right\}$.  
From the coarse-grained Gaussian continuous chain
model\,\cite{matsen2002standard}, 
the single-chain partition function $Q$ and the monomer density
$\rho_\alpha$ can be computed through solving a set of modified
diffusion equations (MDEs) of chain propagators $q_\alpha(\mathbf{r},s)$ and
$q^\dag_\alpha(\bf r, s)$, $s\in I_\alpha$. These chain
propagators satisfy MDEs
\begin{align}
	\frac{\partial}{\partial s}q_\alpha({\bf r}, s) =
	\nabla^2_{\bf r} q_\alpha({\bf r}, s) - \omega_\alpha({\bf
	r})q_\alpha({\bf r}, s), ~~ s\in I_\alpha,
	\label{eqn:qMDE}
	\\
	\frac{\partial}{\partial s}q^\dag_\alpha({\bf r}, s) =
	\nabla^2_{\bf r} q^\dag_\alpha({\bf r}, s) - \omega_\alpha({\bf
	r})q^\dag_\alpha({\bf r}, s), ~~ s\in I_\alpha.
	\label{eqn:qplusMDE}
\end{align}
The initial values of these MDEs are relative to the topological
structure of the polymer
chain\,\cite{fredrickson2006equilibrium}. After solving the MDEs,
we can calculate the single-chain partition function
\begin{align}
	Q[\bm \omega] = \frac{1}{V}\int d{\bf r}\,q_\alpha({\bf r}, s)
	q^\dag_\alpha({\bf r}, 1-s), ~~~ \forall \alpha
	~~\mbox{and}~~  \forall s \in I_\alpha,
	\label{eqn:singlePar}
\end{align}
and the monomer density
\begin{align}
	\rho_\alpha = -V\Big<\frac{\delta\log Q}{\delta \omega_\alpha} \Big>
	= \frac{1}{Q}\int_{I_\alpha} ds\,q_\alpha({\bf r}, s)
	q^\dag_\alpha({\bf r}, 1-s).
	\label{eqn:density}
\end{align}

\subsection{SCFT equations}
\label{subsec:SCFTeqns}

The direct method to investigate the phase behavior of block
copolymer systems is to compute the partition function
\eqref{eq:FieldBasedPartitionFunctional2}\,\cite{fredrickson2006equilibrium}.
However, it is very
difficult to obtain the analysis solution for such a complicated
path integral. Fortunately, since the integrand is an exponential
of energy functional in the complex plane, the saddle point
(mean-field) approximation can be used to calculate partition
function to high accuracy when concerning the most possibly
configuration and ignoring the fluctuations. In the saddle point
approximation, the equilibrium fields $\mu_+^*$, $\bm\mu^*$
dominate the functional integral
\eqref{eq:FieldBasedPartitionFunctional2}.
The field configurations are obtained by demanding that
\eqref{eq:Hamiltonian2} be stationary with respect to fields
$\mu_+$ and $\bm\mu$, i.e.,
\begin{equation}
\begin{aligned}
	\frac{\delta H(\mu_+, \bm\mu)}{\delta
	\mu_+}\Bigg|_{(\mu_+,~\bm\mu)=(\mu_+^*,~ \bm\mu^*)} &= 0,
	\\
	\frac{\delta H(\mu_+, \bm\mu)}{\delta
	\mu_k}\Bigg|_{(\mu_+,~\bm\mu)=(\mu_+^*,~ \bm\mu^*)} &= 0.
\end{aligned}
	\label{eq:scftDervs}
\end{equation}
Expanding the derivatives, then
\begin{align}
	\frac{\delta H(\mu_+^*, \bm\mu^*)}{\delta \mu_+^*} &=
	\frac{n}{V}\Bigg(\sum_\alpha \rho_\alpha(\bm r\,; [\mu_+^*,
	\bm\mu^*]) -1 \Bigg)=0,
	\label{eqn:derivMu0}
	\\
	\frac{\delta H(\mu_+^*, \bm\mu^*)}{\delta \mu_k^*} &=
	\frac{n}{V}\Bigg(\frac{\mu_k^*}{2\zeta_k N}-\sum_\alpha
	\sigma_{k\alpha} \rho_\alpha(\bm r\,; [\mu_+^*, \bm\mu^*])
	\Bigg)=0, \hspace{2mm}~~~ \zeta_k > 0,
	\label{eqn:derivMu1}
	\\
	\frac{\delta H(\mu_+^*, \bm\mu^*)}{\delta \mu_k^*} &=
	-\frac{n}{V}\Bigg(\frac{\mu_k^*}{2\zeta_k N}+\sum_\alpha
	\sigma_{k\alpha} \rho_\alpha(\bm r\,; [\mu_+^*, \bm\mu^*])
	\Bigg)=0, ~~~ \zeta_k < 0.
	\label{eqn:derivMu2}
\end{align}
It is well known in the polymer physics literature as SCFT
equations. Having obtained the saddle point potentials $\mu_+^*$,
$\bm\mu^*$, one can complete the approximation by imposing 
$Z\approx \exp(-\beta H[\mu_+^*, \bm\mu^*])$, and obtain the
value of energy functional $\beta H[\mu_+^*, \bm\mu^*]=-\ln Z$.
The SCFT equations are a set of highly nonlinear system.
Numerically solutions of the SCFT equations are obtained
using iterative techniques. 
Thus analysing the mathematical structure of the energy
functional is useful for prior to attempting its computation.

\subsection{Analytic structure of the field theory}
\label{subsec:saddle}

For the purpose of implementing the saddle point approximation,
it is important to get some understanding of the saddle point
approximation, thereby examining the analytic structure of the
partition function\,\eqref{eq:FieldBasedPartitionFunctional2} as
well as the energy functional\,\eqref{eq:Hamiltonian2}. 
We consider the following simple integral 
\begin{align}
	Z(\lambda) = \int_C g(z)e^{\lambda h(z)} dz,
	\label{eq:sdaExp}
\end{align}
where $C$ is a contour in the complex plane, and $\lambda$ is
a real positive number. 
$g(z)$ and $h(z)$, which are independent of $\lambda$, are
analytic functions of $z$ in some domain of the complex plane
including $C$. 
As a convenient preliminary, we introduce the functions
$u$ and $v$ corresponding to the real part of
and imaginary part of $h(z)$ by writing 
\begin{align}
	h(z) = u(z)+iv(z).
	\label{}
\end{align}
The problem is to find an asymptotic approximation for
$Z(\lambda)$ in the limit $\lambda\to \infty$.
We now exploit the fact that the contour $C$ in \eqref{eq:sdaExp}
can be deformed, by Cauchy's theorem, into other contour
$C'$ that shares the same end points, and have
\begin{align}
	Z(\lambda) = \int_C g(z)e^{\lambda h(z)} dz
	= \int_{C'} g(z)e^{\lambda h(z)} dz.
	\label{eq:sdaExpC}
\end{align}
Such reexpression of $Z$ is useful only if the integral along
$C'$ is easier than the integral along $C$.
If $v(z)$, the so-called phase, is a constant along contour $C' :
v(z)|_{z\in C'} \equiv v_0$, there are no oscillations from
$e^{i\lambda v(z)}$.  For illustrative purpose, let us use
Cartesian coordinates $x$, $y$ and write
\begin{equation}
 z = z(t) = x(t) + iy(t), \hspace{2mm} t \in [a, b],
 \label{eq:paramterContour}
\end{equation}
possibly with $a = -\infty$ or $b = + \infty$ or both, is a
contour in the complex $z$-plane.
Then we have
\begin{equation}
  Z(\lambda) = e^{i\lambda v_0} \int_a^b e^{\lambda u(x(t),
  y(t))} u(t) dt + i e^{i \lambda v_0} \int_a^b e^{\lambda u(x(t), y(t))}v(t) \mathrm d t,
  \label{eq:InteConstPhase}
\end{equation}
where $u(t)$ is the real part and $v(t)$ is the imaginary part of
the function $g(z(t))z'(t)$. 
We deform the path $C$ into $C'$ so that it passes
through the saddle point $z_0$ (i.e., $h'(z_0)=0$), where $u$ has
its maximum value. 
By the Laplace method\,\cite{miller2006applied}, the major
contributions to the integrals \eqref{eq:InteConstPhase} come
from the neighbourhoods of the points of maximum $u(x,y)$ in the
range of integration, as $\lambda\to +\infty$.
Such a procedure is the basis
of the steepest descent (or saddle point) method for the
asymptotic analysis of integrals such as
Eqn.\,\eqref{eq:sdaExp}\,\cite{miller2006applied}.

We now return to $Z(\lambda)$ given by \eqref{eq:sdaExp}
with the given conditions on $g$, $h$, $\lambda$ to further
observe the saddle point approximation. 
Firstly we show that $z_0$ is indeed a saddle point of
$u$ (and of $v$). A relative maximum of $u$ is given by
$z_0=x_0+y_0$. Since $u$ and $v$ are the real and imaginary
parts of $h(z)$, an analytic function of $z$, they satisfy the
Cauchy-Riemann equations which are 
\begin{align}
	u_x = v_y, ~~~ u_y = -v_x.
	\label{eq:CReqn}
\end{align}
Thus
\begin{align}
	h'(z) = u_x + iv_x = u_x - i u_y = 0 ~~
	\mathrm{when}~~ z=z_0.
	\label{}
\end{align}
From \eqref{eq:CReqn}, $u$ and $v$ are harmonic functions
satisfying Laplace's equation 
\begin{align}
	\Delta u = 0, ~~~ \Delta v=0,
	\label{}
\end{align}
where $\Delta$ is the Laplacian operator 
$\partial^2/\partial x^2 +\partial^2/\partial y^2$.
By the maximum modulus theorem, $u$ and $v$ can not have a
maximum or a minimum in the domain of analyticity of $h(z)$. Thus
the point $z_0$ is a saddle point of $u$ and of $v$.
Subsequently we point out that this specific path $C'$ through
the point of maximum $e^{\lambda u}$ will be a path of steepest
descents. 
Assume that $\bm s = (dx, dy)$ is the tangential vector
along the contour $C'$. Since $dv(x,y)=0$ along $C'$, the
following equation holds
\begin{align}
	v_x dx + v_y dy = 0, ~~~ z\in C'.
	\label{}
\end{align}
By Cauchy-Riemann equation \eqref{eq:CReqn}, we have
\begin{align}
	-u_y dx + u_x dy = 0.
	\label{}
\end{align}
In fact, the vector $\bm\tau = (-u_y, u_x)$ is orthogonal to
the $\nabla u$, i.e., $\bm\tau\cdot\nabla u=0$, thus $\bm s
\parallel \nabla u$. That is for $h'(z)\neq 0$, along $C'$
($C': v(z)|_{z\in C'} = \mathrm{const}$), $u$ drops off on
either side of its maximum as rapidly as possible. What is more,
when we concern with saddle points of order one, that is
$h'(z_0)=0$, $h^{''}(z_0)\neq 0$,
there exists another contour
$C^{''}$ passing through $z_0$ along which $v(x,y)$ is also a
constant\,\cite{bender1999advanced}. The contours
$C'$ and $C^{''}$ are orthogonal to each other along which
$u$ changes as rapidly as possible. 
Assume that
$\bm\tau \parallel \bm \gamma = (-dy, dx)$, and 
\begin{align}
	\frac{\partial^2 u(x,y)}{\partial \bm s^2} + \frac{\partial^2
	u(x,y)}{\partial \bm \gamma^2} = 
	\frac{\partial^2 u(x,y)}{\partial x^2} +
	\frac{\partial^2 u(x,y)}{\partial y^2} =0.
	\label{}
\end{align}
As a consequence, at the saddle point $z_0$, 
\begin{equation}
	\frac{\mathrm d^2 u(x,y)}{\mathrm d \bm s^2}\bigg|_{z_0}
	= -\frac{\mathrm d^2 u(x,y)}{\mathrm d \bm
	\gamma^2}\bigg|_{z_0} < 0.
  \label{eq:SaddlePoint}
\end{equation}
Thus the contour $C^{''}$ corresponds not to the steepest descent
path but to the steepest ascent path since the value of $u$ on it
is such that $u(x,y)>u(x_0,y_0)$ except at $z_0$.
The above results are valid for multidimensional
cases\,\cite{miller2006applied}. There are also a few researches
for infinite-dimensional cases\,\cite{ellis1982asymptotic,
ellis1981asymptotic, ellis1982laplace}.

We now return to the field model of multicomponent block
copolymer system which is a functional integral. 
Strictly speaking, it is a quite challenge to find the constant
phase contour $C'$ for such a complicated infinite dimensional integral. 
However, for numerical purpose the fields are represented in a
discrete way and the functional integral is approximated by
multiply integrals of finite, but large, dimension.
Thus the above heuristic discussions about the saddle point
approximation are useful, at least conceptually, to determine the
qualitative location and orientation of a saddle point in the
complex plane. 

The first thing is to verify the possibility of
applying saddle point approximation to model
\eqref{eq:FieldBasedPartitionFunctional2}. 
It is easily verified that the statistical weight $\exp(-H[\mu_+,
\bm\mu])$ is analytic functional of fields $\mu_+(\bm r)$,
$\bm\mu(\bm r)$. Thus, the deformation of integration path is
conceptually possible.
The original path of integration of
the fields $\mu_+(\bm r)$, $\bm\mu(\bm r)$ in the partition function of
\eqref{eq:FieldBasedPartitionFunctional2} is along the real axis.
Nevertheless, for analytic integrands $\exp(-H[\mu_+, \bm\mu])$, it
is useful to deform the integration path
onto a constant phase that passes through one or more saddle
points $H[\mu_+^*, \bm\mu^*]$ in the multi-dimensional complex
plane. The phase factor $\exp(i H_I)$ is a constant along
such a contour, so that oscillations of the integrand are eliminated.
A functional integral deformed onto a constant phase contour is
dominated by saddle point field configurations, $H[\mu_+^*,
\bm\mu^*]$, which correspond to mean-field solutions. 
On the constant phase contour, $H_R$, the real part of $H[\mu_+,
\bm\mu]$, has local minima at the
saddle points. The extent to which one of these saddle point
field configurations dominates the integral depends on the value
of a ``Ginzburg parameter'' analogous to $\lambda$ in the above
discussion. Indeed, it is straightforward to show that the
coordination number $C=(n/V) R_g^3$ is the relevant
parameter\,\cite{fredrickson2006equilibrium}, where $R_g$ denotes
the radius of gyration of a polymer. It follows that the
mean-field approximation becomes asymptotically exact for
$C\to+\infty$. Since polymers are asymptotically ideal in the
melt with $R_g = b(N/6)^{1/2}$, it follows that $C\sim(\rho_0
b^3)N^{1/2}$. Thus the saddle point approximation is accurate 
for concentrated solutions or melts of high-molecular-weight polymers.  

The requirement that the energy functional $H[\mu_+^*, \bm\mu^*]$
be real implies that $\mu_+^*$, $\mu_k^*$ ($\eta_k<0$) must be
imaginary, and $\mu_k^*$ ($\eta_k>0$) must be real, respectively.
It is often convenient to compute a purely imaginary saddle point
by a relaxation scheme along the imaginary axis, which is a
search direction that is orthogonal to the physical path of
integration. With such a scheme, it is important to recognize that 
it is more than likely a descent-ascent direction for the real
part of the energy functional $H[\mu_+, \bm\mu]$. Thus finding
a saddle point of $H[\mu_+, \bm\mu]$, corresponding to the SCFT
solution, shall be maximized with respect to the fields $\mu_+$
and $\mu_k$ ($\zeta_k < 0$), and minimized with respect to the
``exchange chemical potentials'' $\mu_k$ ($\zeta_k > 0$). 
Thus the gradient-based iterative methods, based on the
orientation of saddle points of the energy functional, can be
devised to solve the SCFT equations
\eqref{eqn:derivMu0}-\eqref{eqn:derivMu2}.

The SCFT equations \eqref{eqn:derivMu0}-\eqref{eqn:derivMu2} have
multiply solutions corresponding to more than one saddle point,
which present the stable and metastable phases for polymer
fluids. Inhomogeneous saddle points of interest normally require
numerical iterative techniques. Within the mean-field theory, 
only in the vicinity of a saddle point, the orientation of the
saddle point can be determined, as discussed above. Thus initial
configurations shall be close to the desired saddle point in numerical
computations. 

The above analysis will be helpful to translate the saddle point problem 
into a extreme problem for multicomponent block copolymer system.
For binary component incompressible polymer systems, to find
saddle points, one should seek a local maximum along the exchange
chemical potential $\mu_-$, and a local minimum along the
pressure potential $\mu_+$\,\cite{ganesan2001field,
fredrickson2002field, fredrickson2006equilibrium}. The latter
field is produced by the incompressibility constraint.
Then the energy functional $H[\mu_+, \mu_-]$ can be computed by
$H[\mu_+^*(\mu_-), \mu_-]=H[\mu_-]$ when satisfying the
incompressibility condition.
That is the ``partial mean-field
approximation''\,\cite{reister2001spinodal, duchs2003fluctuation}
which translates the saddle point problem into a minimum value problem.  
The ``partial mean-field approximation'' is useful for 
the nucleation theory of ordered
phases\,\cite{cheng2010nucleation} and the mesoscopic
dynamics\,\cite{fraaije1993dynamic, fraaije1997dynamic}.
It is different from the multicomponent incompressible
polymer systems. In these models, the incompressibility condition
$\delta[\hat\rho(\bm r)-1]$ is also the unique constraint
condition, however, the nature of the saddle point may come
from the non-bond Flory-Huggins interaction parameters
$\chi_{\alpha\beta}$.
When all parameters $\zeta_k$, the combination of
$\chi_{\alpha\beta}$, are positive, the SCFT solutions are local
minima of $H[\mu_+^*(\bm\mu), \bm\mu]=H[\bm\mu]$. 
Once there exist one or more parameters such that
$\zeta_k<0$, the SCFT solutions are still saddle points
of $H[\mu_+^*(\bm\mu), \bm\mu]$. Thus for multicomponent
polymer systems, the original ``partial mean-field
approximation'' cannot avoid saddle point problem. It requires
the new ``partial mean-field approximation'', 
$H[\mu_+^*(\mu_{k;~\zeta_k >0}),
\mu^*_{k;~\zeta_k<0}(\mu_{k;~\zeta_k >0})]$ to translate the saddle
point problem into a minimum value problem. It follows that the string
method can be applied to study the nucleation of ordered phases
for multicomponent block copolymers.

Though continuous Gaussian chain model is considered here,
extension to more complicated systems, for example rod-coil block
copolymers and liquid crystalline side-chain copolymers with the 
wormlike chain model should be straight forward.
Direct extensions can also be easily applied to grand canonical
ensemble and compressible systems.

\subsection{SCFT of $ABC$ triblock copolymer melt}
\label{subsec:scftABC}

As an example, the SCFT of an incompressible block
copolymer melt with three different chemical species, denoted by
$A$, $B$, $C$, is given in this subsection.
Applying the procedure proposed in
Sec.\,\ref{subsec:statisticalField} to this system, the
interaction energy of the block copolymer melt can be presented
in the form
\begin{equation}
  \begin{split}
     U_1[\bm \Gamma] & = \rho_0\int
	d\mathbf{r}\,[\,\chi_{AB} \hat \rho_A(\bm r) \hat \rho_B(\bm r)
	+ \chi_{AC} \hat \rho_A(\bm r) \hat \rho_C(\bm r) + \chi_{BC}
	\hat \rho_B(\bm r) \hat \rho_C(\bm r)]
	\\
    & = \rho_0\int
	d\mathbf{r}\,\left\{\frac{\Delta}{4 \chi_{AC}} \hat
	\rho_B^2(\bm r) - \chi_{AC} \left[ \hat \rho_C(\bm r) +
	\frac{\chi_{AB} + \chi_{AC} - \chi_{BC}}{2 \chi_{AC}} \hat
	\rho_B(\bm r)\right]^2 \right.
	\\
	&\hspace{7cm} \left.+ \chi_{AB} \hat \rho_B (\bm r) + \chi_{AC} \hat \rho_C(\bm
	r)\right\},
  \end{split}
  \label{eq:InteractingEnergy}
\end{equation}
where
\begin{align}
  \Delta = \chi_{AB}^2+ \chi_{AC}^2+ \chi_{BC}^2
  -2\chi_{AB}\chi_{AC}-2\chi_{AB}\chi_{BC}-2\chi_{AC}\chi_{BC}.
  \label{eqn:triblock:oscmft:delta}
\end{align}
Omitting the linear terms, the non-bonded interaction energy
becomes 
\begin{equation}
     U_1[\bm \Gamma] 
     = \rho_0\int
	d\mathbf{r}\,\left\{-\zeta_1 \hat\rho_1^2 - \zeta_2
	\hat\rho_2^2 \right\}, ~~~~~~~\hat\rho_k = \sum_\alpha
	\sigma_{k\alpha} \hat{\rho}_\alpha, ~~~ k = 1,2.
  \label{eq:InteractingEnergy2}
\end{equation}
In this expression, 
\begin{equation}
\zeta_1 = \frac{-\Delta}{4\chi_{AC}}, \hspace{8mm} \zeta_2 = \chi_{AC},
\end{equation}
and the coefficients $\sigma_{k\alpha}$ that have been satisfied
the orthogonal relation $\hat\rho_+ \cdot \hat\rho_k = 0$ are
\begin{equation}
\begin{aligned}
&\sigma_{1A} = \frac{1}{3}, \hspace{11mm}
\sigma_{1B} = -\frac{2}{3}, \hspace{11.5mm}
\sigma_{1C} = \frac{1}{3},
\\
&\sigma_{2A} = \frac{1+\alpha}{3},\hspace{5mm}
\sigma_{2B} = \frac{1-2\alpha}{3}, \hspace{6mm}
\sigma_{2C} = \frac{\alpha-2}{3}, \hspace{6mm}
\alpha = \frac{\chi_{AC}+\chi_{AB}-\chi_{BC}}{2\chi_{AC}}.
\end{aligned}
\end{equation}

As discussed in the above subsection, the value of $\zeta_k$
influences the saddle point character of the SCFT. In the following,
we will analyse the expression of the SCFT for different $\zeta_k$.
Without loss of generality, we will assume
$\chi_{\alpha\beta}>0$, $\alpha, \beta \in \{A,B,C\}$.
The similar case of $\chi_{\alpha\beta}<0$ or
$\chi_{\alpha\beta}\neq0$ can be discussed in the same way.

If $\Delta \neq 0$, the field-based energy functional
can be written as
\begin{align}
  \label{eqn:triblock:oscmft:hamilton}
	H &= \frac{n}{V}\int
	d\mathbf{r}\,\Big(\frac{1}{4N\zeta_1}\mu_1^2 +
	\frac{1}{4N\zeta_2}\mu_2^2 - \mu_+ \Big)
	- n \log Q[\omega_A, \omega_B, \omega_C],
	\\
\omega_\alpha &= \mu_+ - \sigma_{1\alpha}\mu_1 -
\sigma_{2\alpha}\mu_2, ~~~ \alpha \in \{A,B,C\}.
	\label{eqn:triblock:oscft:w}
\end{align}
The corresponding SCFT equations are
\begin{align}
	& \rho_A+\rho_B+\rho_C-1=0,
	\label{eqn:triblock:oscmft:muPlus}
    \\
	& \frac{1}{2N\zeta_1}\mu_1 - \sigma_{1A}\rho_A -
		\sigma_{1B}\rho_B - \sigma_{1C}\rho_C  = 0,
	\label{eqn:triblock:oscmft:muMinus1}
    \\
	& \frac{1}{2N\zeta_2}\mu_2 - \sigma_{2A}\rho_A -
		\sigma_{2B}\rho_B - \sigma_{2C}\rho_C  = 0.
	\label{eqn:triblock:oscmft:muMinus2}
\end{align}
When $\Delta > 0$, the equilibrium solutions of the 
energy functional\,(\ref{eqn:triblock:oscmft:hamilton}) are to be
maximized with respect to the potential field $\mu_+$ and
$\mu_1$, and minimized with respect to another potential field
$\mu_2$. When $\Delta < 0$, the equilibrium states of the
energy functional\,(\ref{eqn:triblock:oscmft:hamilton}) are
the maxima along the potential field $\mu_+$,
and minima along other potential fields $\mu_1$, $\mu_2$.
When $\Delta = 0$, the ``exchange chemical field'' $\mu_1$
disappears. The field-based energy functional
degenerates to 
\begin{align}
  \label{eqn:triblock:hamilton:0}
	H &= \frac{n}{V}\int
	d\mathbf{r}\,\Big(
	\frac{1}{4N\zeta_2}\mu_2^2 - \mu_+ \Big)
	- n \log Q[\omega_A, \omega_B, \omega_C],
	\\
\omega_\alpha &= \mu_+ -
\sigma_{2\alpha}\mu_2, ~~~ \alpha \in \{A,B,C\}.
	\label{eqn:triblock:oscft:w}
\end{align}
It follows that the SCFT equations are
\begin{align}
	& \rho_A+\rho_B+\rho_C-1=0,
	\label{eqn:triblock:muPlus}
    \\
	& \frac{1}{2N\zeta_2}\mu_2 - \sigma_{2A}\rho_A -
		\sigma_{2B}\rho_B - \sigma_{2C}\rho_C  = 0.
	\label{eqn:triblock:muMinus2}
\end{align}
In this case, the physical solutions of the
incompressible block copolymer system are saddle points in
which the energy
functional\,(\ref{eqn:triblock:hamilton:0}) shall be
maximized along field $\mu_+$, and
minimized along field $\mu_2$.

The above SCFT system is closed by combining with the
single-chain partition function $Q$ of (\ref{eqn:singlePar}),
density $\rho_\alpha$ (\ref{eqn:density}) and the forward and
backward propagators of
MDEs \eqref{eqn:qMDE}-\eqref{eqn:qplusMDE}.
The topological structure of the block copolymer
chain determines the initial values of
MDEs\,\cite{matsen2002standard, fredrickson2006equilibrium}.
For the $ABC$ linear triblock copolymer chain,
\begin{equation}
\begin{aligned}
	&	q_A(\bm r, 0) = 1, \hspace{1.55cm} q_C^\dag(\bm r, 0) = 1,
	\\
&	q_B(\bm r, 0) = q_A(\bm r, f_A), ~~~~~~ q_B^\dag(\bm r, 0) = q_C^\dag(\bm r, f_C),
	\\
&	q_C(\bm r, 0) = q_B(\bm r, f_B), ~~~~~~ q_A^\dag(\bm r, 0) = q_B^\dag(\bm r, f_B).
\end{aligned}
	\label{}
\end{equation}
For the $ABC$ star triblock copolymer chain,
\begin{equation}
	q_\alpha(\mathbf{r}, 0) = 1, \hspace{1.cm}
	q_\alpha^\dag(\mathbf{r}, 0) = q_\beta(\mathbf{r},
	f_\beta) q_\gamma(\mathbf{r}, f_\gamma),
	\label{}
\end{equation}
where $(\alpha\beta\gamma)\in\{(ABC),(BCA),(CAB)\}$.

\section{Gradient-based Iterative Methods}
\label{sec:method}

In the preceding discussions, we have been analysed the analytic
structure of the SCFT energy functional of multicomponent polymer
systems, which enables us to cast light on the orientation of the
saddle points corresponding to the SCFT solutions.
Thus the existing gradient-based methods 
can be planted into multicomponent polymer systems.
Essentially the SCFT equations are a set of
highly nonlinear equations with multi-solutions and
multi-parameters. It requires iterative techniques
to solve the complicated system.
For given parameters $\chi_{\alpha\beta}$,
$f_\alpha$, and fixed the calculation box, the specific iteration
processes for the self-consistent systems are given in the
following:
\begin{description}
  \item[Step 1] Give the initial values of chemical potential fields;
  \item[Step 2] Solve propagators to
	obtain single partition function $Q$ and density
	operators $\rho_\alpha$;
  \item[Step 3] Update potential fields $\mu_+, \bm\mu$ through 
	  iterative methods;
  \item[Step 4] If a given convergent condition is achieved, stop
	  iteration procedure, else goto \textbf{Step 2}.
\end{description}

The numerical search for equilibrium ordered structures depends
on the initial conditions. A series of effective strategies have been
developed for screening initial values to discover ordered
patterns\,\cite{jiang2010spectral, jiang2013discovery, xu2013strategy}.
The subsequently numerical experiments will utilize these
strategies to screen initial configurations. 

Evaluating the energy functional and first derivatives of the
self-consistent field models requires an efficient numerical
method to solve MDEs (\ref{eqn:qMDE}) and (\ref{eqn:qplusMDE})
for propagators. The forth-order accurate Adams-Bashford scheme
with the pseudospectral technique\,\cite{cochran2006stability} is
our method of choice,
\begin{align}
  \frac{25}{12}q^{j+1}-4q^{j}+3q^{j-1}-\frac{4}{3}q^{j-2}+\frac{1}{4}q^{j-3}=
	\Delta
	s[\nabla^2q^{j+1}-w(4q^j-6q^{j-1}+4q^{j-2}-q^{j-3})].
	\label{eqn:adams}
\end{align}
The initial values required to apply the multi-step scheme of
(\ref{eqn:adams}) are obtained by using second-order
operator-splitting scheme\,\cite{rasmussen2002improved} and
Richardson's extrapolation.
The expressions for the derivatives of the energy functional are
linearly related to density operators, which are, in turn, nonlinearly and
nonlocally related to the propagators $q$ and $q^\dag$ by
expressions of (\ref{eqn:qMDE}) and (\ref{eqn:qplusMDE}). These
expressions are most conveniently evaluated in real space at the
collocation points of the basis functions. In particular, the
composite Simpson's rule with forth-order truncation
error (see Ref.\,\cite{press2007numerical}, p.134) is utilized in
our numerical implementation.

Based on the proposed SCFT, we have known that at saddle points
$H[\mu_+, \bm \mu]$ achieves its local maxima
along $\mu_+$ and $\mu_k$ ($\zeta_k < 0$), and local minima along
$\mu_k$ ($\zeta > 0$). Thus gradient-based iterative methods,
such as the explicit forward Euler method, 
hybrid method\,\cite{liang2013efficient},
and the semi-implicit method\,\cite{ceniceros2004numerical}, 
can be extended to update chemical potential fields of the SCFT
energy functional for multicomponent block polymer systems.

\textbf{Explicit forward Euler method (EFE)}.
A natural approach is to introduce a fictitious ``time'' variable
$t$ and relax the gradient of $H[\mu_+, \bm \mu]$, which yields
\begin{equation}
\begin{aligned}
	\frac{\partial}{\partial t}\mu_+(\bm r, t) &=
	\frac{\delta H[\mu_+, \bm\mu]}{\delta \mu(\bm r, t)},
	\\
	\frac{\partial}{\partial t}\mu_k(\bm r, t) &=
	\frac{\delta H[\mu_+, \bm\mu]}{\delta \mu_k(\bm r, t)},
	 ~~~~~~~ \zeta_k < 0,
	\\
	\frac{\partial}{\partial t}\mu_k(\bm r, t) &= -
	\frac{\delta H[\mu_+, \bm\mu]}{\delta \mu_k(\bm r, t)},
	 ~~~~ \zeta_k > 0.
\end{aligned}
	\label{eqn:iterDGM}
\end{equation}
A simple and useful explicit scheme is the forward Euler formula
\begin{equation}
\begin{aligned}
	\mu_+^{j+1}(\bm r) &= \mu_+^j(\bm r) + \lambda_+ \frac{\delta
	H[\mu_+^j, \bm \mu^j]}{\delta
	\mu_+^j(\bm r)},
	\\
	\mu_k^{j+1}(\bm r) &= \mu_k^j(\bm r) + \lambda_k
	\frac{\delta H[\mu_+^j, \bm \mu^j]}{\delta
	\mu_k^j(\bm r)}, ~~~~ \zeta_k < 0,
	\\
	\mu_k^{j+1}(\bm r) &= \mu_k^j(\bm r) - \lambda_k
	\frac{\delta H[\mu_+^j, \bm \mu^j]}{\delta
	\mu_k^j(\bm r)}, ~~~~ \zeta_k > 0.
\end{aligned}
	\label{eqn:iterEFE}
\end{equation}
where iterative steps $\lambda_+, \lambda_k > 0$,
These derivatives are computed by Eqn.\,\eqref{eq:scftDervs}.

\textbf{Hybrid conjugate gradient method (HCG)}.
The second explicit method is the hybrid scheme,
firstly developed by Liang \textit{et
al.}\,\cite{liang2013efficient} for flexible-semiflexible diblock
copolymer system. The conjugate gradient method is normally unstable in
the SCFT simulations\,\cite{ceniceros2004numerical}.
However, the hybrid method combines the robustness of the
steepest descent method with the efficiency of the conjugate
gradient method. When minimizing an objective functional $F[\mu]$,
the hybrid conjugate gradient method reads:
\begin{center}
\begin{algorithmic}
	\STATE Let $i = 0$, give initial values $\mu_0$,
	\\calculate $g_0= \delta F[\mu]/\delta \mu$, $d_0 = -g_0$
	\WHILE{$\delta F[\mu]/\delta \mu \neq 0$}
	\STATE $i\Leftarrow i+1$
	\STATE choose $\lambda_i$ to minimize
	$F[\mu_i+\lambda_i d_i]$
	\STATE $\mu_{i+1} = \mu_i +\lambda_i d_i$
	\STATE $g_{i+1} = \delta F[\mu_{i+1}]/\delta \mu$
	\STATE $\gamma_{i+1}=(g_{i+1}^{T}g_{i+1})/(g_{i}^{T}g_{i})$
	\STATE $d_{i+1} = -g_{i+1}+\alpha_H\gamma_{i+1}d_i$
	\ENDWHILE
\end{algorithmic}
\end{center}
In the algorithm, $\alpha_H$ is a hybrid factor to adjust the conjugate
gradient direction. If $\alpha_H = 0$, the hybrid scheme degenerates
to steepest descent scheme. If $\alpha_H=1$, it is the
conjugate gradient scheme. By choosing $0 < \alpha_H < 1$, the hybrid method
takes the advantage of the robustness of steepest descent method
and the efficiency of conjugate gradient scheme. The method is
not only applied to find the minima of an objective function, but
also can be suitable for accomplishing the maxima of a given
functional $F[\mu]$ through first variation $\delta F[\mu]/\delta
\mu$ multiplied by $-1$. Therefore, the chemical potential fields
$\mu_+$ and $\bm \mu$ can be all updated simultaneously by the
HCG method in the light of the orientation of the saddle point of
the SCFT.

\textbf{Semi-implicit method (SIM)}.
The third iterative method is a semi-implicit scheme,
originally developed by Ceniceros and
Fredrickson\,\cite{ceniceros2004numerical} for incompressible
systems of homopolymer system, diblock copolymer melt and binary
homopolymer blend.
The pressure potential is updated using the approximated
second-order variation information of $\frac{\delta^2 H}{\delta
\mu_+(\bm r)\delta \mu_+(\bm r')}$, while the exchange chemical
potential field is updated with the new field $\mu_+$
\,\cite{ceniceros2004numerical}. The approximated second-order
variation of the energy functional $H$ with respect to
$\mu_+$ is obtained by the asymptotic expansion (or named as random
phase approximation expansion\,\cite{fredrickson2006equilibrium}).
Here we can extend the idea to multicomponent polymer
systems based on the developed SCFT framework. 
The expression of $\frac{\delta^2 H}{\delta \mu_+(\bm r)\delta
\mu_+(\bm r')}$ has been given in Appendix \ref{app:semi}.
As the Ref.\,\cite{ceniceros2004numerical} recommended,
we propose the semi-implicit method:
\begin{align}
	\hat{\mu}^{j+1}_{+}(\mathbf{k}) &=
	\hat{\mu}^j_{+}(\mathbf{k})^j+
	\Bigg(\frac{1}{\lambda_+}+\hat{G}(\bm k)\Bigg)^{-1}
  \widehat{\Bigg(\frac{\delta H[\mu_+^j, \bm \mu^j]}{\delta
  \mu_+^j}\Bigg)}(\bm k), ~~~ \mbox{for all spectral elements }
  \hat{\mu}_{+}(\bm k),
  \label{eqn:simnm:mu0}
  \\
	\mu_k^{j+1}(\bm r) &= \mu_k^j(\bm r) + \lambda_k
	\frac{\delta H[\mu_+^{j+1}, \bm \mu^j]}{\delta
	\mu_k^j(\bm r)}, ~~~~ \zeta_k < 0,
	\\
	\mu_k^{j+1}(\bm r) &= \mu_k^j(\bm r) - \lambda_k
	\frac{\delta H[\mu_+^{j+1}, \bm \mu^j]}{\delta
	\mu_k^j(\bm r)}, ~~~~ \zeta_k > 0,
\end{align}
where the caret denotes Fourier transform.
The iterative steps $\lambda_+, \lambda_k$ are
positive numbers.
In the scheme, pressure potential field $\mu_+$ can be updated
in Fourier-space conveniently by applying an FFT-inverse FFT
pair with optimal computational complexity, whilst
the chemical potential fields $\mu_k$ can be updated
in both real-space and Fourier space.
We note that the second solutions of the
MDEs\,(\ref{eqn:qMDE})-(\ref{eqn:qplusMDE}) are
required before updating potentials $\mu_k$ in order to evaluate
$\delta H[\mu_+^{j+1}, \bm \mu^j] / \delta \mu_k^j(\bm r)$.
This approximately doubles the computational cost per time over
the explicit schemes. The step
lengths $\lambda_k$ are allowed to be larger in
updating remaining chemical potential fields $\mu_k$ with the
characteristic of implicity.

\section{Numerical Results}
\label{sec:results}

We will apply the developed SCFT and the extended gradient-based
methods, namely EFE, HCG and SIM schemes, to $ABC$ star
triblock copolymer melts by a series of numerical experiments in
three spatial dimensions. 
All the methods considered here have been implemented in C++ language.
The cubic computational box with edge length $D$ and periodic
boundary conditions is imposed on all numerical examples.
Fourier transforms in pseudo-spectral method are
calculated by using the FFTW package\,\cite{frigo1998fftw}.
The numerical experiments were run in the same computer, a Intel(R)
Xeon(R) CPU E5520 @2.27 GH, memory - 24 GB, under
linux. The total chain contour is discretized into $10^2$ points
(i.e., $\Delta s =0.01$) in computing propagators $q$ and
$q^\dag$ with forth-order Adams-Bashford method of
(\ref{eqn:adams}). The uniform
discretization with $24\times 24\times24$ plane waves for
three-dimensional problems is used in the SCFT calculations.
To measure the error, we use the $l^{\infty}$ norm
\begin{align}
	E = \max_{\gamma}\Big\{ E_{\gamma} \Big\},
	\hspace{0.5cm} 	\mbox{where}~~~~
	E_{\gamma} = \left|\left|\frac{\delta H}{\delta
	\mu_\gamma}\right|\right|_{l^{\infty}},
	\hspace{0.5cm} \gamma = +, 1 ,2.
	\label{}
\end{align}
From our experience, this kind of convergence criterion will
produce more precise results than the relative error of the
energy functional (between two consecutive iterations) at the
same error tolerance.
The step sizes, $\lambda_+$, $\lambda_{k}$, and the hybrid
factor $\alpha_H$ in HCG are always selected as large as the
stability of the methods permits and so that the error is
decreased the fastest.

The value of $\Delta$ expressed by
(\ref{eqn:triblock:oscmft:delta}) for the systems with three
different chemical species influences the saddle point character
of the physical relevant solutions as discussed in
Sec.\ref{subsec:scftABC}. Thus, numerical experiments for
$\Delta < 0$, $\Delta = 0$, and $\Delta > 0$ will be implemented
in the following on a case-by-case basis.

We first consider the case of $\Delta<0$ which is also the common
used parameter range in SCFT simulations for three different chemical
species systems\,\cite{tang2004morphology, tyler2007scft,
liu2012theoretical, tang2004star, xu2013strategy, li2009theory}.
By the analysis of the saddle point properties of SCFT solutions,
to achieve the equilibrium state of $ABC$ triblock
copolymer model, the energy functional shall be minimized with respect
to the potential fields $\mu_1$ and $\mu_2$, and maximized with
respect to pressure potential field $\mu_+$.
In the first numerical experiment the Flory-Huggins interaction
parameters are $[\chi_{BC}N,\chi_{AC}N,\chi_{AB}N]=[30,45,30]$.
We take into account an asymmetric case $[f_A,
f_B,f_C]=[0.20, 0.70, 0.10]$. 
The A15 spherical pattern of core-shell version is the target
pattern. We input the space symmetric group
$Pm\bar{3}n$\,\cite{jiang2013discovery} on the chemical potential
field $\mu_1$ as initial values. A cube with $D=7.5$ is
selected as computational box.

\begin{figure}[!htbp]
	\centering
	\subfigure[]{
	  \includegraphics[scale=0.4]{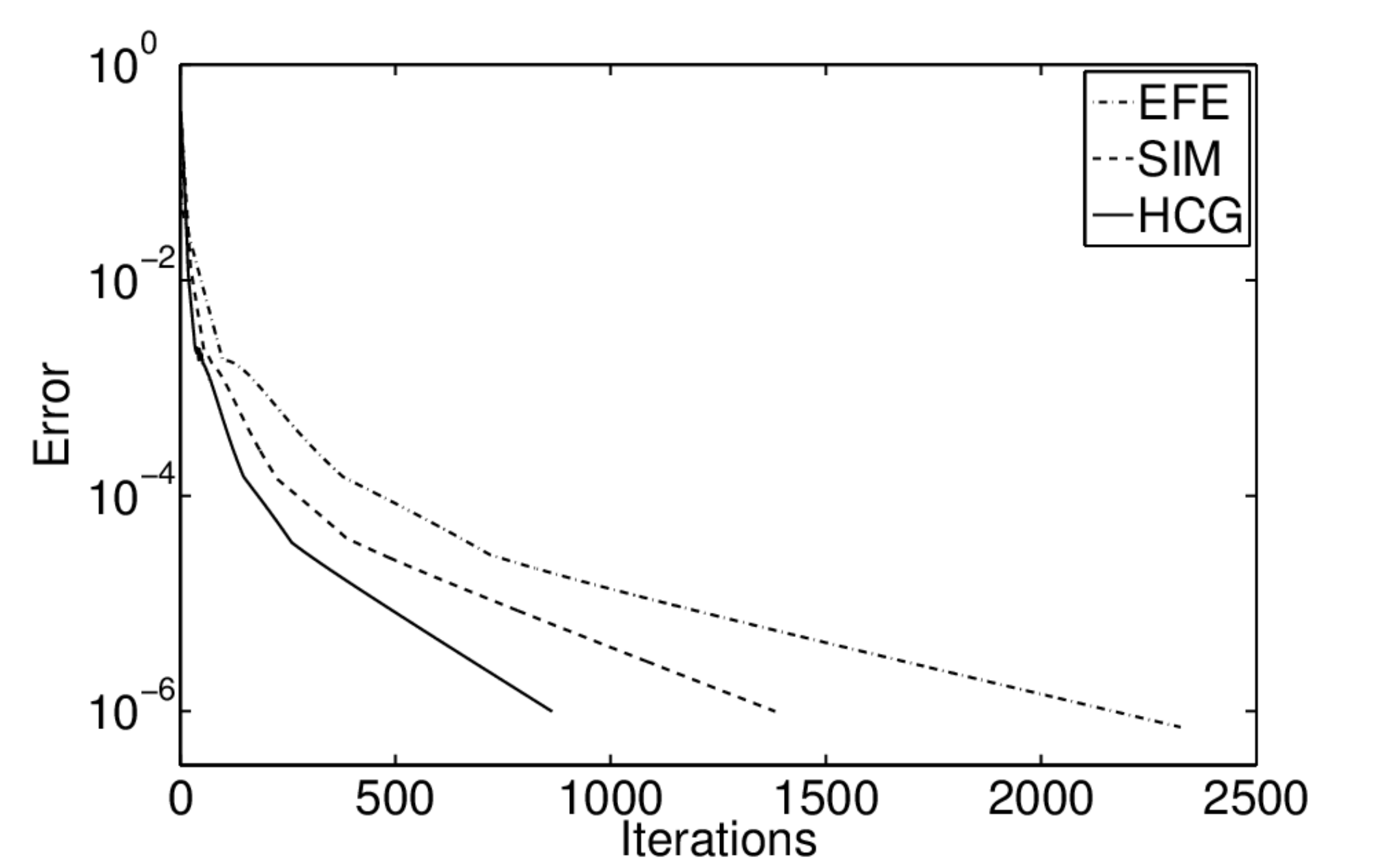}
	\label{subfig:star:err:csA15}
	  }
	\subfigure[]{
	  \includegraphics[width=5.55cm]{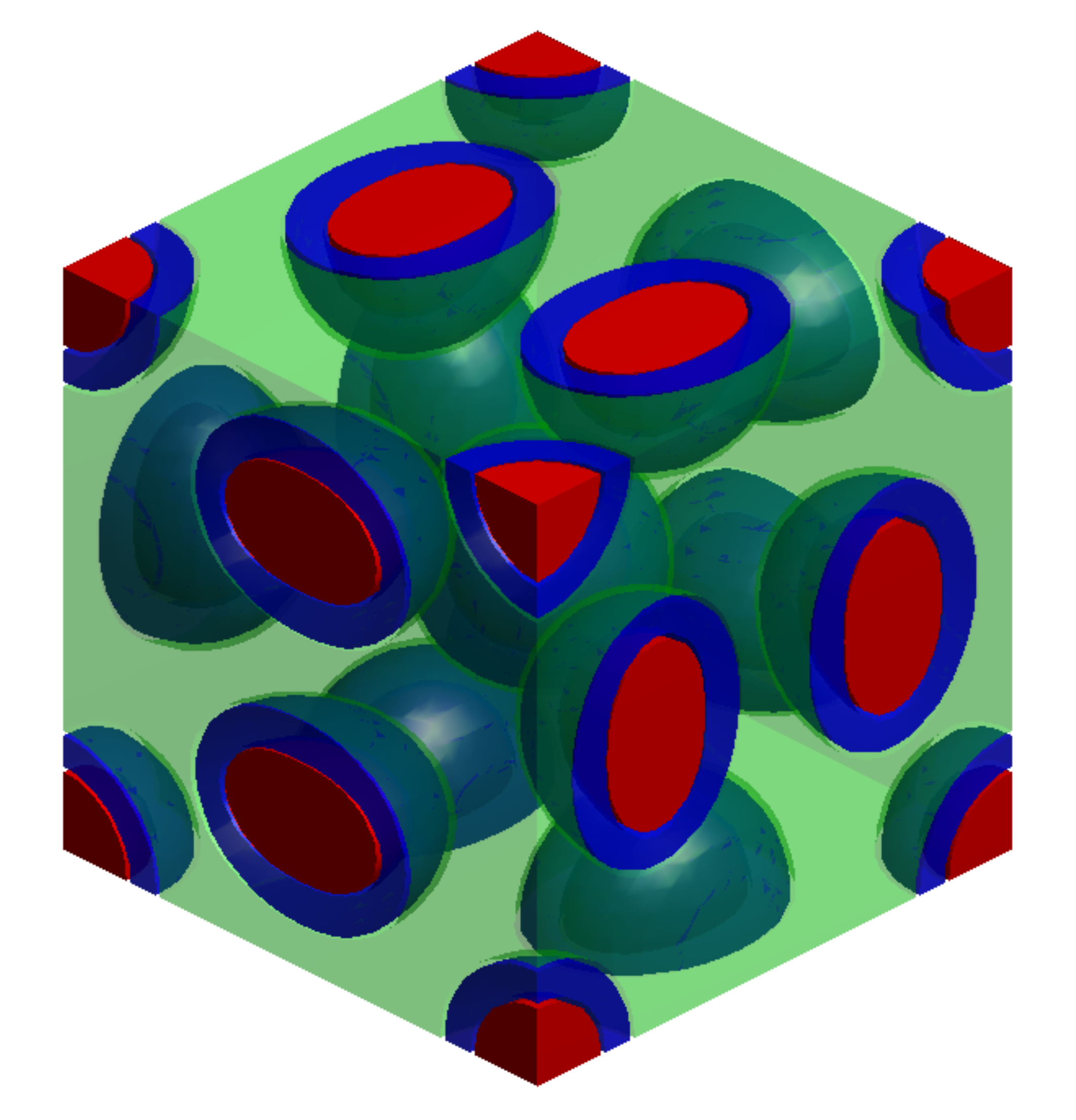}
	\label{subfig:star:phi:csA15}
	}
	\caption{$ABC$ star triblock copolymer melt model:
	$[\chi_{BC}N,\chi_{AC}N,\chi_{AB}N]=[30,45,30]$, $[f_A,
	f_B,f_C]=[0.20, 0.70, 0.10]$, $D=7.5$. (a) Comparison of the EFE
	scheme (dashed-dotted curve), the SIS scheme (dashed curve)
	and the HCG scheme (continuous curve). (b) Core-shell A15 pattern.
	Monomers A, B, and C are denoted by red, green, and blue colors.}
\label{fig:star:csA15}
\end{figure}

The performance of three iterative methods, namely the error
plotted against the number of iterations, is demonstrated in
Fig.\,\ref{subfig:star:err:csA15}.
The detail is presented in Tab.\,\ref{tab:star:csA15}.
The iterations of the SIM scheme multiplied by two illustrate
the feature of semi-implicity that approximately double the
computational costs against the explicit schemes.
Meanwhile Fig\,\ref{subfig:star:phi:csA15} shows
the final solution of core-shell A15.
These numerical results demonstrate that the gradient-based
iterative schemes can reach the saddle points along the analysed
directions.
In particular, the HCG method takes $383$ iterations to reduce
the error to $O(10^{-6})$, SIM scheme costs $1382$ iterations,
and EFE approach spends $2324$ iterations. In the special case
the HCG scheme has the fastest rate of convergence. 
\begin{table}[!htbp]
\caption{\label{tab:star:csA15} $ABC$ star triblock copolymer
melt model: $[\chi_{BC}N,\chi_{AC}N,\chi_{AB}N]=[30,45,30]$, $[f_A,
	f_B,f_C]=[0.20, 0.70, 0.10]$, $D=7.5$.
	Comparison of different iterative methods for spherical phase
	of A15 in core shell version.}
\begin{center}
\begin{tabular}{||c||c|c||c|c||c|c||}
\hline
 & \multicolumn{2}{c||}{EFE} &\multicolumn{2}{c||}{SIM} &\multicolumn{2}{c||}{HCG}
\\ \hline \hline
Parameters & \multicolumn{2}{c||}{$[\lambda_+, \lambda_1, \lambda_2]=[3, 3, 3]$}
 &\multicolumn{2}{c||}{$[\lambda_+, \lambda_1, \lambda_2]=[5, 5, 5]$}
 &\multicolumn{2}{c||}{\makecell{ $[\alpha_{H, +}, \alpha_{H,1}, \alpha_{H,2}]=[0.5, 0.5, 0.5]$ \\
 $[\lambda_+, \lambda_1, \lambda_2]=[4.0, 4.0, 4.0]$}}
\\ \hline
Error
& Iterations & CPU time (sec)
& Iterations & CPU time (sec)
& Iterations & CPU time (sec)
 \\\hline
$E=10^{-6}$
& 2324 & 1022
& 1382$(\times 2)$ & 1234
& 864 & 383
 \\\hline
\end{tabular}
\end{center}
\end{table}

The second case considered here is $\Delta = 0$ for which
$[\chi_{BC}N,\chi_{AC}N,\chi_{AB}N] = [15, 15, 60]$. 
In this case, as Sec.\,\ref{subsec:scftABC} analysed,
the chemical potential field of $\mu_1$ disappears. To achieve a
saddle point, one shall maximize the energy functional along
$\mu_+$, while minimize along $\mu_2$.
$Fm\bar{3}m$ space group symmetry is added into the
initial condition of chemical potential $\mu_2$ to compute the
objective phase of core-shell spheres in face-centered cubic (FCC) lattice.
The volume fractions are $[f_A, f_B,f_C] = [0.20, 0.63, 0.17]$, and $D=4$.

\begin{figure}[!htbp]
	\centering
	\subfigure[]{
	  \includegraphics[scale=0.4]{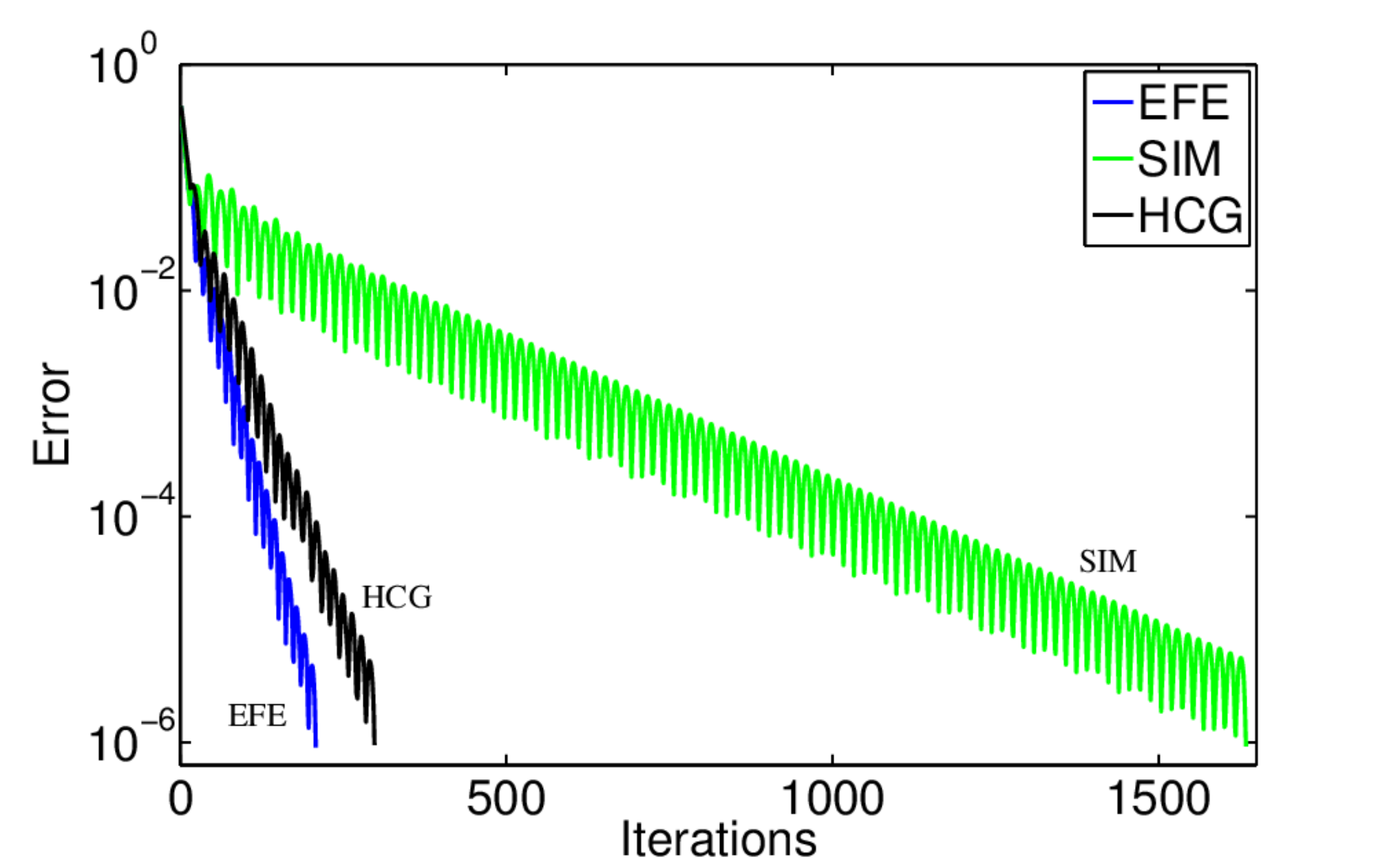}
	\label{subfig:star:err:csFcc}
	  }
	\subfigure[]{
	  \includegraphics[width=5.55cm]{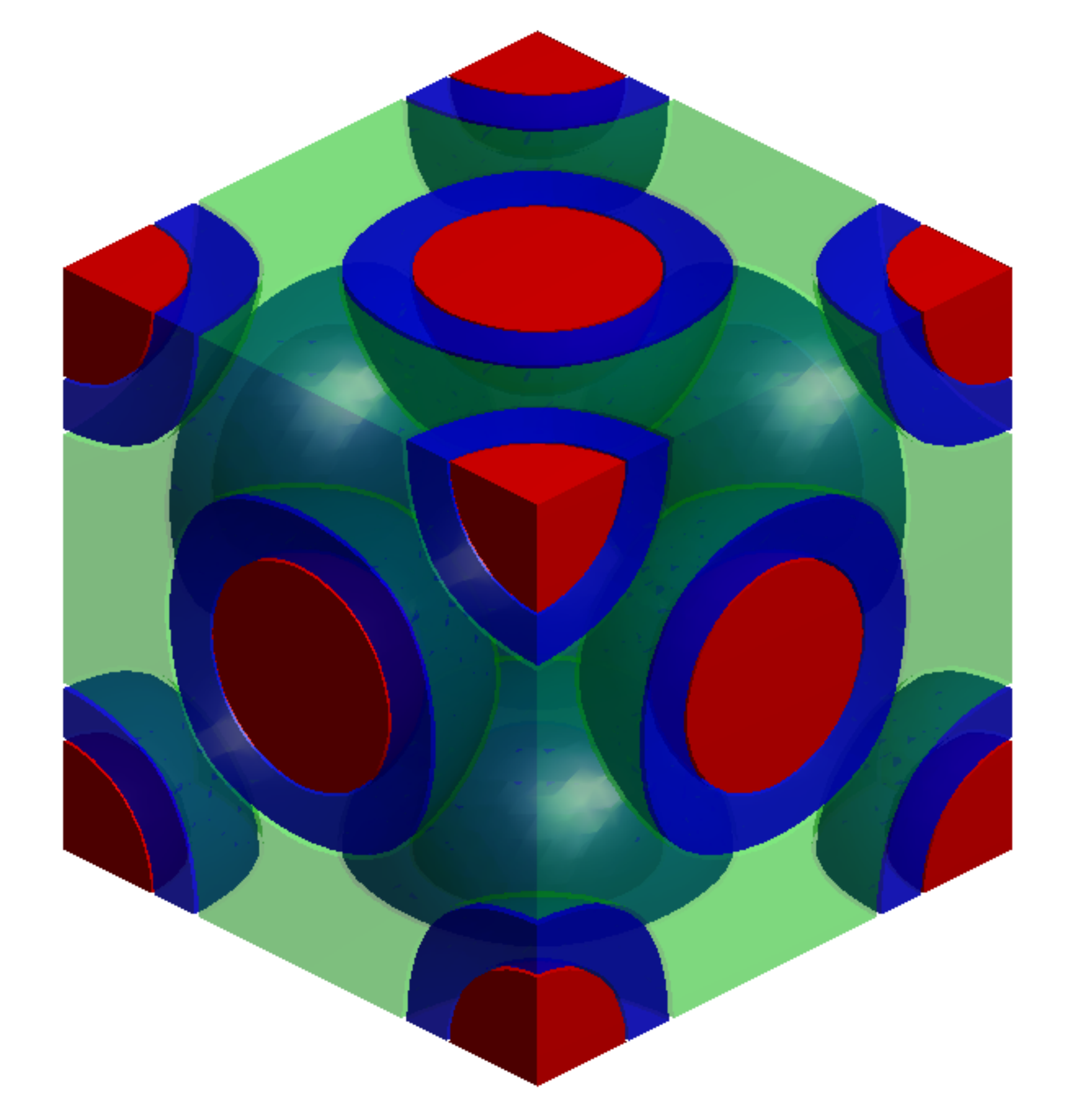}
	\label{subfig:star:phi:csFcc}
	}
	\caption{ABC star triblock copolymer melt model:
	$[\chi_{BC}N,\chi_{AC}N,\chi_{AB}N]=[15,15,60]$, $[f_A,
	f_B,f_C]=[0.20, 0.63, 0.17]$, $D=4$. (a)
	Comparison of the EFE scheme (dashed-dotted curve), the SIS
	scheme (dashed curve) and the HCG scheme (continuous curve).
	(b) Core-shell spheres in FCC lattice.
	Monomers A, B, and C are denoted by red, green, and blue colors.}
\label{fig:star:csFcc}
\end{figure}

Fig.\,\ref{subfig:star:err:csFcc} compares the behavior of the
error for the explicit forward Euler,
semi-implicit, and hybrid conjugate gradient methods. 
In this numerical example, the error of these iterative methods
has an oscillatory behavior, however, the residuals still become
small against the iterations.
The detailed performance of these methods at accuracy of
$10^{-6}$ is listed in Tab.\,\ref{tab:ABCstar:csFcc}.
The final morphology of FCC phase in core-shell version is shown
in Fig\,\ref{subfig:star:phi:csFcc}.
In this case three gradient-based iterative method can obtain the
equilibrium ordered pattern in the light of the new proposed SCFT.
Among these methods, the EFE method has the fewest number of iterators to
reduce the error of $O(10^{-6})$. In particular, the EFE method
is superior to the SIM and for an error of $10^{-6}$ over $7.8$ times.
\begin{table}[!htbp]
\caption{\label{tab:ABCstar:csFcc} $ABC$ star triblock copolymer
melt model: $[\chi_{BC}N,\chi_{AC}N,\chi_{AB}N]=[15, 15, 60]$, $[f_A,
	f_B,f_C]=[0.20, 0.63, 0.17]$, $D=4$.
	Comparison of different iterative methods for FCC spherical
	phase in core-shell version.}
\begin{center}
\begin{tabular}{||c||c|c||c|c||c|c||}
\hline
 & \multicolumn{2}{c||}{EFE} &\multicolumn{2}{c||}{SIM} &\multicolumn{2}{c||}{HCG}
\\ \hline \hline
Parameters & \multicolumn{2}{c||}{$[\lambda_+, \lambda_2]=[4, 4]$}
 &\multicolumn{2}{c||}{$[\lambda_+, \lambda_2]=[6, 6]$}
 &\multicolumn{2}{c||}{\makecell{ $[\alpha_{H, +},
 \alpha_{H,2}]=[0.4, 0.4]$ \\
 $[\lambda_+, \lambda_2]=[2, 2]$}}
\\ \hline
Error
& Iterations & CPU time (sec)
& Iterations & CPU time (sec)
& Iterations & CPU time (sec)
 \\\hline
$E=10^{-6}$
& 208  & 78
& 1634$(\times 2)$ & 1233
& 298  & 112
 \\\hline
\end{tabular}
\end{center}
\end{table}

As the last case we take into account $\Delta > 0$
for $ABC$ star triblock copolymers where the
SCFT solutions become index-2 saddle
points. The equilibrium states of this model shall
be updating up the gradients in $\mu_1$ and $\mu_+$
coordinates and down the gradient in $\mu_2$ to generate the
chemical potentials.
The selection of the parameters of this case is
$[\chi_{BC}N,\chi_{AC}N,\chi_{AB}N] = [52, 12, 12]$.
$[f_A, f_B,f_C] = [0.05, 0.25, 0.70]$.
The initial values of tetragonally ordered cylinder are used for
the chemical potential $\mu_1$. The domain size is $D=7$. 

\begin{figure}[!htbp]
	\centering
	\subfigure[]{
	  \includegraphics[scale=0.4]{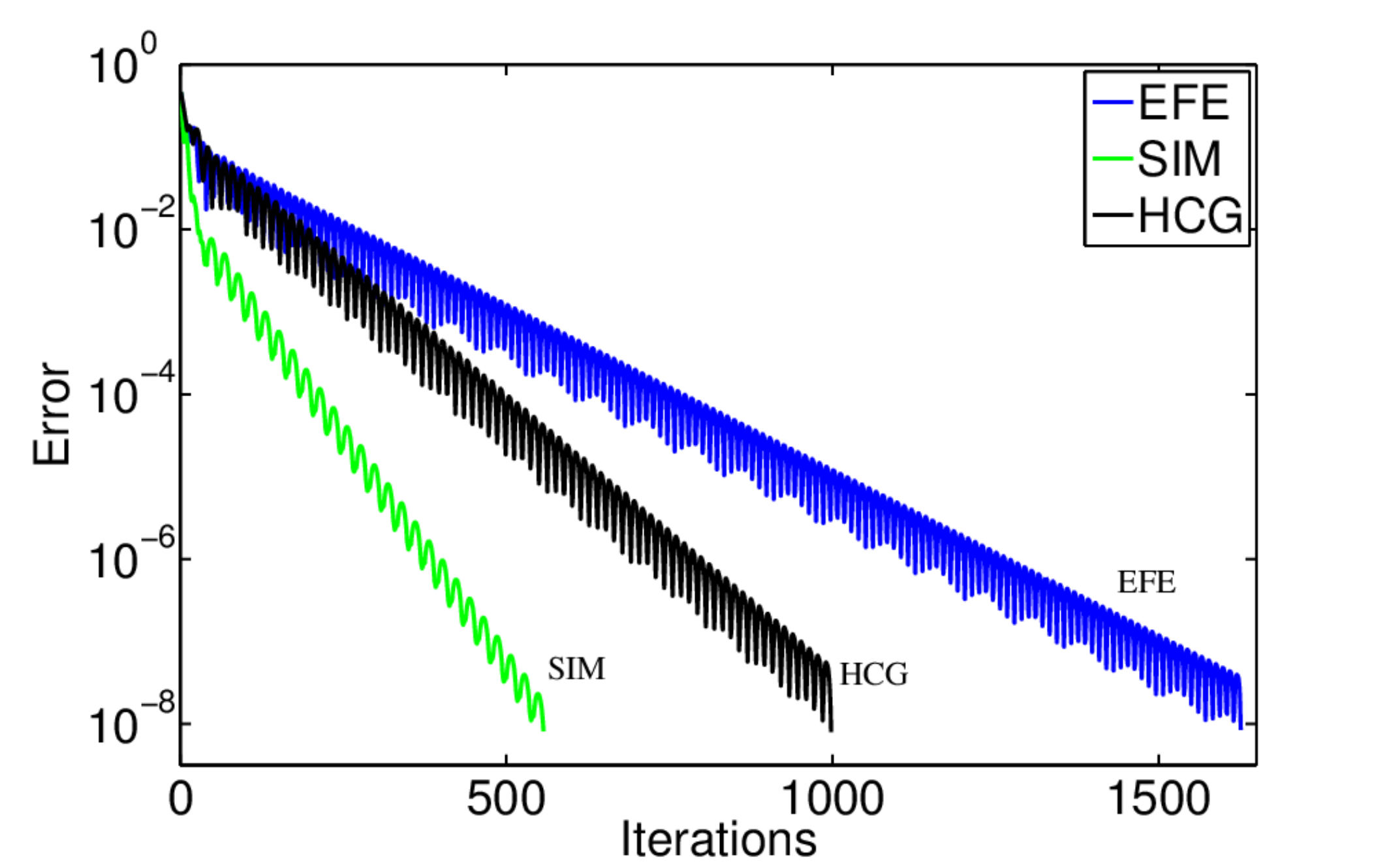}
	\label{subfig:star:err:CoC}
	  }
	\subfigure[]{
	  \includegraphics[width=5.55cm]{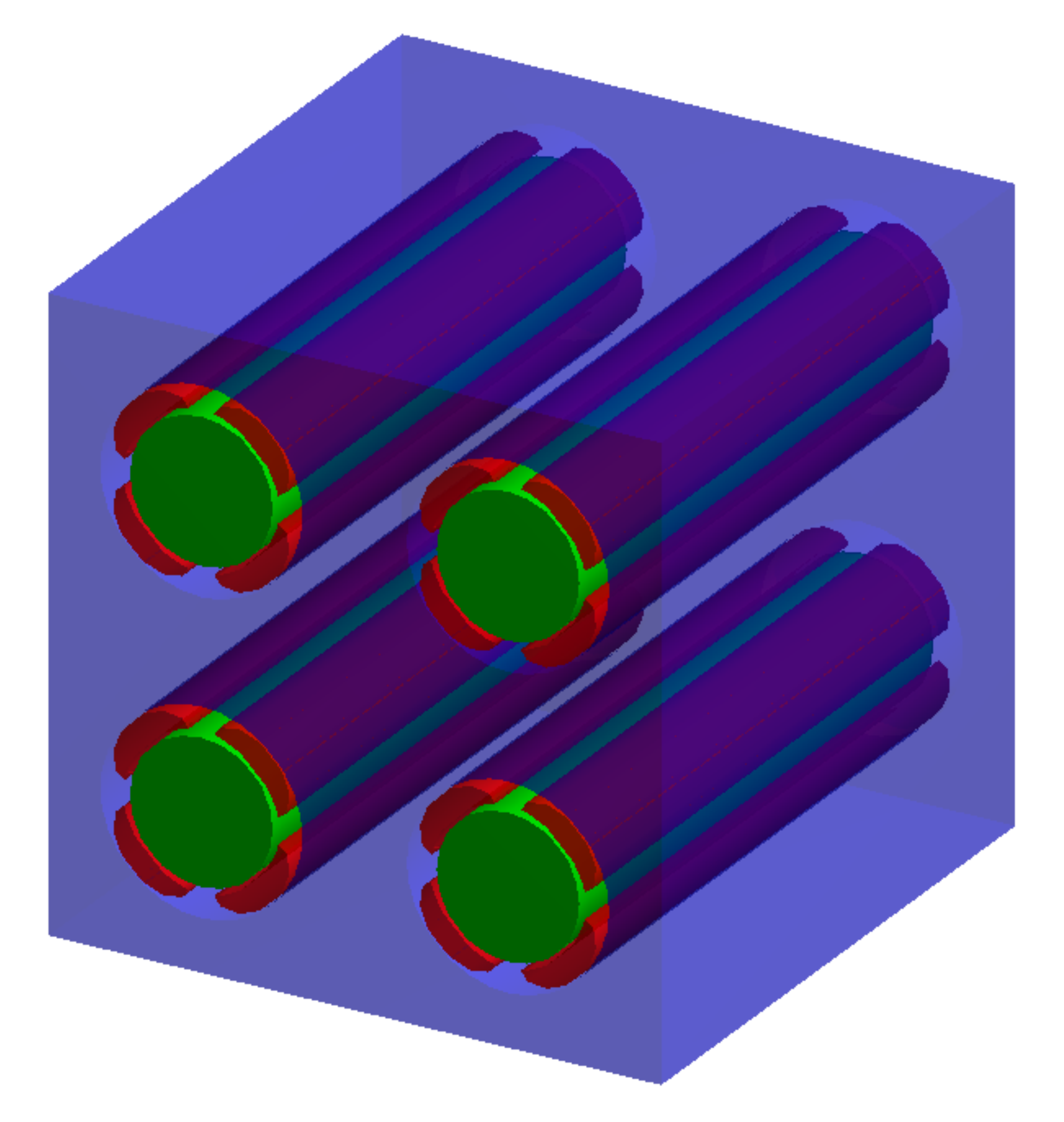}
	\label{subfig:star:phi:CoC}
	}
	\caption{$ABC$ star triblock copolymer melt model:
	$[\chi_{BC}N,\chi_{AC}N,\chi_{AB}N]=[52,12,12]$, $[f_A,
	f_B,f_C]=[0.05, 0.25, 0.70]$, $D=7$. (a)
	Comparison of the EFE scheme (dashed-dotted curve), the SIS
	scheme (dashed curve) and the HCG scheme (continuous curve).
	(b) the morphology of CoC pattern.
	Monomers A, B, and C are denoted by red, green, and blue colors.}
\label{fig:star:CoC}
\end{figure}

Fig.\,\ref{subfig:star:err:CoC} demonstrates the behavior of the
error of the three iterative schemes. 
The final morphology of cylinders on cylinder (CoC) is shown in
Fig\,\ref{subfig:star:phi:CoC}.
For this pattern, the error of these iterative methods shows an
oscillatory behavior, however, the trend of the residual always
declines against iterations.
Tab.\,\ref{tab:ABCstar:CoC} presents a comparsion of their
performance at $O(10^{-8})$. 
The SIM and HCG methods show better numerical performance than
EFE. Moreover, the SIM sheme has the fastest
convergent rate, but spends more computational cost than explicit
hybrid conjugate method because of the extra computational burden
of the semi-implicit character.
The EFE method reaches an error of $O(10^{-8})$ after $611$
seconds ($1626$ iterators), while SIM scheme (or HCG)
accomplishes the same accuracy in just $408$ seconds ($371$
seconds), saving nearly half of CPU time costs compared with the EFE. 
\begin{table}[!htbp]
\caption{\label{tab:ABCstar:CoC} $ABC$ star triblock copolymer
melt model: $[\chi_{BC}N,\chi_{AC}N,\chi_{AB}N]=[52,12,12]$, $[f_A,
	f_B,f_C]=[0.05, 0.25, 0.70]$, $D=7$.
	Comparison of different iterative methods for CoC phase.}
\begin{center}
\begin{tabular}{||c||c|c||c|c||c|c||}
\hline
 & \multicolumn{2}{c||}{EFE} &\multicolumn{2}{c||}{SIM} &\multicolumn{2}{c||}{HCG}
\\ \hline \hline
Parameters & \multicolumn{2}{c||}{$[\lambda_+, \lambda_1, \lambda_2]=[3, 3, 3]$}
 &\multicolumn{2}{c||}{$[\lambda_+, \lambda_1, \lambda_2]=[6, 6, 6]$}
 &\multicolumn{2}{c||}{\makecell{ $[\alpha_{H, +}, \alpha_{H,1}, \alpha_{H,2}]=[0.2, 0.2, 0.2]$ \\
 $[\lambda_+, \lambda_1, \lambda_2]=[2, 2, 2]$}}
\\ \hline
Error
& Iterations & CPU time (sec)
& Iterations & CPU time (sec)
& Iterations & CPU time (sec)
 \\\hline
$E=10^{-8}$
& 1626 & 611
& 557$(\times 2)$ & 408
& 998 & 371
 \\\hline
\end{tabular}
\end{center}
\end{table}

\section{Discussion and Conclusions}
\label{sec:conclusion}

We have utilized the Hubbard-Stratonovich transformation
to establish the field-based theory and
analysed the analytic structure of the SCFT energy functional for
incompressible multicomponent block copolymer systems. 
Our analysis makes clear that the saddle point nature of SCFT solutions
resulting from both the incompressibility constraint and the
combination of the Flory-Huggins interaction parameters. 
Except for the incompressibility constraint, the value of 
$\zeta_k$ (the combination of Flory-Huggins parameters) may give
rise to high index saddle point.
When $\zeta_k < 0$, the energy functional should be maximized
with respect to $\mu_k$.  When $\zeta_k > 0$, the energy
functional is to be minimized with respect to $\mu_k$.  When
$\zeta_k = 0$, the chemical potential $\mu_k$ disappears which
reduces the number of SCFT equations.
It is different from the analytic structure of the SCFT energy
functional of two-component polymer systems whose saddle point
character is only caused by the incompressibility condition.
In this paper, we only concerns the incompressible multicomponent
block copolymer systems with flexible macromolecules, however,
the result can be applied to other polymer or copolymer
systems, such as blends, compressible systems and semiflexible
polymers. 

We believe that the analysis of the SCFT energy functional will
be beneficial to many aspects of the theoretical study of
multicomponent polymeric systems, such as the nucleation theory of
ordered phases using the string
method\,\cite{cheng2010nucleation}, the mean-field mesoscopic
dynamics. In this article we just drew our attention to the development of
iterative methods to solve the SCFT equations. Based upon the
analysis of the SCFT energy functional, three gradient-based
iterative method, originally developed for binary-component
polymer systems, have been extended to multicomponent block
copolymer systems, including the EFE, HCG and SIM methods.
As an application, we took $ABC$ star triblock
copolymer melts to demonstrate the numerical behavior
of these numerical methods. Numerical experiments for $\Delta < 0$,
$\Delta = 0$, and $\Delta > 0$ have been implemented, respectively.
Numerical results indicate that these iterative methods can
efficiently capture the equilibrium states along the 
ascent and descent paths as analysed for each case. 
Meanwhile we should point out that recently
extended gradient-based methods are not the most effective way to
solve the SCFT equations. In fact, in our numerical tests, the Anderson
method\,\cite{thompson2004improved}, which is a multi-step
quasi-Newton method for general fixed-point
problems\,\cite{walker2011anderson}, performs better in the $ABC$
star triblock copolymer model.

\vspace{0.5cm} \noindent{\bf Acknowledgements}
The work is supported by the National 
Science Foundation of China (Grant No.~21274005 and 50930003).

\section*{Appendix}

\begin{appendix}

\section{The transformation of interaction energy}
\label{app:interEnergy}
Using the incompressibility condition
$\hat{\rho}_+=\sum_{\alpha} \rho_\alpha=1$,
the interaction energy of (\ref{eq:QuadraticForm})
in terms of congruent transformation becomes
\begin{equation}
\begin{aligned}
	U_1[\mathbf{R}]  &= \rho_0\int
  d\mathbf{r}\,\sum_{\alpha\neq\beta}\chi_{\alpha\beta}\hat{\rho}_{\alpha}(\mathbf{r})\hat{\rho}_{\beta}(\mathbf{r})
  \\
  &= \rho_0\int d\mathbf{r}\,(\hat{\rho}_A, \hat{\rho}_B, \cdots,
  \hat{\rho}_M)
  \left(
  \begin{array}{ccccc}
	0 & \frac{\chi_{AB}}{2} & \frac{\chi_{AC}}{2} &\cdots &
	\frac{\chi_{AM}}{2}
	\\
	\frac{\chi_{AB}}{2} & 0 & \frac{\chi_{BC}}{2} &\cdots &
	\frac{\chi_{BM}}{2}
	\\
	\vdots &  \vdots & \ddots & \ddots&\vdots
	\\
	\frac{\chi_{AM}}{2} &  \frac{\chi_{BM}}{2} &
	\frac{\chi_{CM}}{2} & \cdots & 0
  \end{array}
  \right)
  \left(
  \begin{array}{c}
	\hat{\rho}_A
	\\
	\hat{\rho}_B
	\\
	\vdots
	\\
	\hat{\rho}_M
  \end{array}
  \right)
  \\
  & = \rho_0\int d\mathbf{r}\,\hat{\Psi}
\small{
  \left(
  \begin{array}{ccccc}
	1 & 0 & 0 &\cdots & 0
	\\
	-1 & 1 & 0 &\cdots & 0
	\\
	\vdots &  \vdots& \ddots & \ddots&\vdots
	\\
	-1 & 0 &0 & \cdots & 0
  \end{array}
  \right)
  \left(
  \begin{array}{ccccc}
	0 & \frac{\chi_{AB}}{2} & \frac{\chi_{AC}}{2} &\cdots &
	\frac{\chi_{AM}}{2}
	\\
	\frac{\chi_{AB}}{2} & 0 & \frac{\chi_{BC}}{2} &\cdots &
	\frac{\chi_{BM}}{2}
	\\
	\vdots &  \vdots& \ddots & \ddots&\vdots
	\\
	\frac{\chi_{AM}}{2} &  \frac{\chi_{BM}}{2} &
	\frac{\chi_{CM}}{2} & \cdots & 0
  \end{array}
  \right)
  }
  \small{
  \left(
  \begin{array}{ccccc}
	1 & -1 & -1 &\cdots & -1
	\\
	0 & 1 & 0 & \cdots & 0
	\\
	\vdots &  \vdots& \ddots & \ddots&\vdots
	\\
	0 & 0 &0 & \cdots & 0
  \end{array}
  \right)
  }
  \hat{\Psi}^{T}
  \\
  & = \rho_0\int d\mathbf{r}\,\hat{\Psi}
  \cdot S_1 \cdot
  \hat{\Psi}^{T}
  \\
  & = \rho_0\int d\mathbf{r}\, \hat{\Psi}
  \cdot S_2 \cdot
  \hat{\Psi}^{T}
  +C(\chi_{\alpha\beta}, \rho_0, N_\alpha, V),
  \\
  & = -\rho_0 \int d\mathbf{r}\,
  \sum\limits_{k=1}^{M-1}\zeta_k
  \left(\sum_{\alpha=1}^{M}\sigma_{k\alpha}\hat{\rho}_\alpha\right)^2 +
  C(\chi_{\alpha\beta}, \rho_0, N_\alpha, V),
  \\
  & = -\rho_0 \int d\mathbf{r}\,
  \sum\limits_{k=1}^{M-1}\zeta_k \hat{\rho}_k^2 +
  C(\chi_{\alpha\beta}, \rho_0, N_\alpha, V)
  \\
  &\stackrel{\Delta}{=} G\left[ \hat \rho_B(\bm r), \hat
  \rho_C(\bm r) \dots \right] + L\left[ \hat \rho_B(\bm r), \hat
  \rho_C(\bm r) \dots \right],
  \label{eqn:oscft:interactionEnergy}
\end{aligned}
\end{equation}
where 
\begin{align*}
	&\small{
	S_1 = 
  \left(
  \begin{array}{ccccc}
	0 & \frac{\chi_{AB}}{2} & \frac{\chi_{AC}}{2} &\cdots &
	\frac{\chi_{AM}}{2}
	\\
	\frac{\chi_{AB}}{2} & -\chi_{AB} & \cdots & \cdots &
	\\
	\frac{\chi_{AC}}{2}&  & -\chi_{AC} & (s_{ij})  & \vdots
	\\
	\vdots & (s_{ji})   & \ddots  & \ddots&  \vdots
	\\
	\frac{\chi_{AM}}{2}&  &\cdots & \cdots  & -\chi_{AM}
  \end{array}
  \right),
  ~~
	S_2 = 
  \left(
  \begin{array}{ccccc}
	0 & 0 & 0 &\cdots & 0
	\\
	0 & -\chi_{AB} &  &  \cdots &
	\\
	0&  & -\chi_{AC} &  (s_{ij})  & \vdots
	\\
	\vdots & (s_{ji}) & \ddots & \ddots&  \vdots
	\\
	0&  & \cdots &  \cdots & -\chi_{AM}
  \end{array}
  \right)
  }~,
  \\
  & s_{ij} = s_{ji} =
  \frac{\chi_{ij}-\chi_{1i}-\chi_{1j}}{2},
  ~~ j>i>2.
\end{align*}
$\hat{\Psi}=(\hat{\rho}_+, \hat{\rho}_B, \cdots,\hat{\rho}_M)$,
$\zeta_k$ are the compositions of $\chi_{\alpha\beta}$,
$\hat{\rho}_k=\sum_{\alpha=1}^M
\sigma_{k\alpha}\hat{\rho}_\alpha$, $\sigma_{k\alpha}$ are
constants. The constant term $C$ just represents a constant shift
in energy functional and dose not influence the configuration of
the system. We here utilize the incompressibility condition to
eliminate the $\hat{\rho}_A$ in the transformation of interaction
energy. There are also other options, such as eliminating
$\hat\rho_\alpha$ ($\alpha\neq A$) to translate $U_1[\bm R]$. The
difference is only a constant, but it also does not influence the
configuration of the system.

\section{Gaussian functional integrals}
\label{app:gauss}

The following Gaussian functional integrals using in the
Hubbard-Stratonovich transformation can be also found in
Refs.\,\cite{chaikin1995principles, fredrickson2006equilibrium}.
\begin{align}
  \nonumber
  \frac{\int Df\,\exp[-(1/2)\int\,dx\int\,dx'\,f(x)A(x,x')f(x') + \int\,dx\, J(x)f(x)]}
  {\int\,Df \exp[-(1/2)\int\,dx\int\,dx'\,f(x)A(x,x')J(x')]}
  \\
  \label{eqn:gaussPathIntMin}
  =\exp\left(\frac{1}{2}\int\,dx\int\,dx'\,J(x)A^{-1}(x,x')J(x')\right),
  \\ \nonumber
  \frac{\int Df\,\exp[-(1/2)\int\,dx\int\,dx'\,f(x)A(x,x')f(x') + i\int\,dx\, J(x)f(x)]}
  {\int\,Df \exp[-(1/2)\int\,dx\int\,dx'\,f(x)A(x,x')J(x')]}
  \\
  \label{eqn:gaussPathIntMax}
  =\exp\left(-\frac{1}{2}\int\,dx\int\,dx'\,J(x)A^{-1}(x,x')J(x')\right),
\end{align}
where $A(x, x')$ is assumed to be real, symmetric, and positive
definite. The functional inverse of $A$, $A^{-1}$, is defined by
\begin{align}
	\int dx'\,A(x,x')A^{-1}(x', x'') = \delta(x-x'').
	\label{}
\end{align}
When these formulas are applied to interacting particle models in
classical statistical physics. $J$ represents a microscopic
density operator, and $A$ is a pair potential function. The
function $f$ is an auxiliary potential that serves to decouple
particle-particle interactions.

\section{The second-order variation of $H$ with respect to $\mu_+$} 
\label{app:semi}

In this Appendix, we use the random phase approximated
expansion\,\cite{fredrickson2006equilibrium} to obtain the
approximate expression of the second-order variation of the SCFT
energy functional $H$ with respect to $\mu_+$.
We firstly consider the $ABC$ triblock linear copolymer systems.
A particularly useful perturbation
expansion can be derived when the potential field
$\omega_\alpha$, where $\alpha = A,B,C$ has inhomogeneities that
are weak in amplitude. We expand potential fields $\omega_\alpha$ as
\begin{align}
\omega_\alpha(\mathbf{r}) = \omega_{0,\alpha}+\varepsilon
\omega_{1,\alpha},
~ ~
\omega_{0,\alpha} =
\frac{1}{V}\int\,d\mathbf{r}\,\omega_\alpha(\mathbf{r}),
\end{align}
$\varepsilon$ is a small parameter $\varepsilon \ll 1$.
Let's translate propagator $q(\mathbf{r},s)$ into
\begin{align}
	q(\mathbf{r},s) =
	p(\mathbf{r},s)\exp\Big(-\int_0^s\,d\tau\,\omega_0(\mathbf{r},s)\Big),
\end{align}
which leads to
\begin{align}
	 \frac{\partial
	p}{\partial s} = \nabla^2
	p(\mathbf{r},s)-\varepsilon\omega_{1}p(\mathbf{r},s),
	~ ~  p(\mathbf{r},0) = 1.
	\label{inhomoPDE}
\end{align}
A weak inhomogeneity expansion can be developed by assuming that
$p(\mathbf{r},s)$ can be expressed as
\begin{align}
p(\mathbf{r},s) \sim \sum\limits_{j=1}^{\infty}\varepsilon^j
p^{(j)}(\mathbf{r},s).
\label{weakInhomo}
\end{align}
Using these expressions, single chain partition function
$Q$ (see Eqn.(\ref{eqn:singlePar})) can be written as
\begin{align}
	Q = \hat{q}(\mathbf{0},1) =
	\exp\Big(-\int_0^1\,d\tau\,\omega_0(\mathbf{r},\tau)\Big)
	\Big[{p}^{(0)}(\mathbf{r},1)+\varepsilon{p}^{(1)}(\mathbf{r},1)+\varepsilon^2{p}^{(2)}(\mathbf{r},1)+
	o(\varepsilon^3) \Big].
\end{align}
In particular,
\begin{align}
	p^{(0)}(\mathbf{r},1) &= 1,
	\\
	\nonumber
	\hat{p}^{(1)}(\mathbf{k},s) &=
	\frac{\hat{\omega}_{1,A}(\mathbf{k})}{\mathbf{k}^2}(1-e^{\mathbf{k}^2
	f_A})e^{-\mathbf{k}^2 s}
	+\frac{\hat{\omega}_{1,B}(\mathbf{k})}{\mathbf{k}^2}(1-e^{\mathbf{k}^2
	f_B})e^{-\mathbf{k}^2(s-f_A)}
	\\ & \hspace{0.3cm}
	+\frac{\hat{\omega}_{1,C}(\mathbf{k})}{\mathbf{k}^2}(1-e^{\mathbf{k}^2(s-f_A-f_B)})e^{-\mathbf{k}^2(s-f_A-f_B)},
	\\ \nonumber
	\hat{p}^{(2)}(\mathbf{0},1) &=
	\frac{1}{2}\int\,d\mathbf{k}_1\,\hat{\omega}_{1,C}(-\mathbf{k}_1)\hat{\omega}_{1,C}(\mathbf{k}_1)\hat{g}_{CC}
	+\frac{1}{2}\int\,d\mathbf{k}_1\,\hat{\omega}_{1,C}(-\mathbf{k}_1)\hat{\omega}_{1,A}(\mathbf{k}_1)\hat{g}_{AC}
	\\ \nonumber  & \hspace{0.3cm}
	+\frac{1}{2}\int\,d\mathbf{k}_1\,\hat{\omega}_{1,C}(-\mathbf{k}_1)\hat{\omega}_{1,B}(\mathbf{k}_1)\hat{g}_{BC}
+ \frac{1}{2}\int\,d\mathbf{k}_1
\,\hat{\omega}_{1,B}(-\mathbf{k}_1)\hat{\omega}_{1,B}(\mathbf{k}_1)\hat{g}_{BB}
	\\ & \hspace{0.3cm}
	+\frac{1}{2}\int\,d\mathbf{k}_1
	\,\hat{\omega}_{1,A}(-\mathbf{k}_1)\hat{\omega}_{1,A}(\mathbf{k}_1)\hat{g}_{AA}
	 +\frac{1}{2}\int\,d\mathbf{k}_1
	\,\hat{\omega}_{1,B}(-\mathbf{k}_1)\hat{\omega}_{1,A}(\mathbf{k}_1)\hat{g}_{AB},
\end{align}
where
\begin{equation}
	\begin{aligned}
	& \hat{g}(\mathbf{k}, s) =
	\frac{2}{\mathbf{k}^2}\Big(e^{-\mathbf{k}s}+\mathbf{k}s-1
	\Big),
	\\
	& \hat{g}_{AA} = \hat{g}(\mathbf{k}^2,f_A),
	\\ &
	\hat{g}_{BB} = \hat{g}(\mathbf{k}^2,f_B),
	\\ & \hat{g}_{AB} =
	\hat{g}(\mathbf{k}^2,f_A+f_B)-
	\hat{g}(\mathbf{k}^2,f_B)-\hat{g}(\mathbf{k}^2,f_A),
	\\ &
	\hat{g}_{CC} = \hat{g}(\mathbf{k}^2,f_C),
	\\ & \hat{g}_{AC} =
	\hat{g}(\mathbf{k}^2,1)- \hat{g}(\mathbf{k}^2,f_A+f_B)-
	\hat{g}(\mathbf{k}^2,1-f_A)+ \hat{g}(\mathbf{k}^2,f_B),
	\\ &
	\hat{g}_{BC} = \hat{g}(\mathbf{k}^2,1-f_A)-
	\hat{g}(\mathbf{k}^2,f_B)-\hat{g}(\mathbf{k}^2,f_C).
	\end{aligned}
\end{equation}
Therefore,
\footnotesize{
\begin{eqnarray} 
	\nonumber 
	Q \sim
\exp\Big(-\int_0^1\,d\tau\,\omega_0(\mathbf{r},\tau)\Big)\Big[
1-\varepsilon \hat{\omega}_{1,A}(0) f_A -\varepsilon
\hat{\omega}_{1,B}(0)f_B -\varepsilon \hat{\omega}_{1,C}(0)f_C +
\frac{\varepsilon^2}{2}\int\,
d\mathbf{k}\,\hat{\omega}_{1,C}(-\mathbf{k})\hat{\omega}_{1,C}(\mathbf{k})\hat{g}_{CC}
\\ \nonumber + \frac{\varepsilon^2}{2}\int\,
d\mathbf{k}\,\hat{\omega}_{1,C}(-\mathbf{k})\hat{\omega}_{1,A}(\mathbf{k})\hat{g}_{AC}
+ \frac{\varepsilon^2}{2}\int\,
d\mathbf{k}\,\hat{\omega}_{1,C}(-\mathbf{k})\hat{\omega}_{1,B}(\mathbf{k})\hat{g}_{BC}
+ \frac{\varepsilon^2}{2}\int\,
d\mathbf{k}\,\hat{\omega}_{1,B}(-\mathbf{k})\hat{\omega}_{1,B}(\mathbf{k})\hat{g}_{BB}
\\
\nonumber
+ \frac{\varepsilon^2}{2}\int\,
d\mathbf{k}\,\hat{\omega}_{1,A}(-\mathbf{k})\hat{\omega}_{1,A}(\mathbf{k})\hat{g}_{AA}
 + \frac{\varepsilon^2}{2}\int\,
d\mathbf{k}\,\hat{\omega}_{1,B}(-\mathbf{k})\hat{\omega}_{1,A}(\mathbf{k})\hat{g}_{AB}
\Big].
\end{eqnarray}
}

\normalsize{
Note that $\hat{\omega}_{\alpha,1}(0) = 0$, we have
}
\begin{align}
	\nonumber &
	\frac{\partial\log Q}{\partial
	\hat{\omega}_A(-\mathbf{k})}
	=\frac{1}{Q} \frac{\partial Q}{\partial
	\hat{\omega}_A(-\mathbf{k})}
	\thickapprox
	\hat{\mu}_+(\frac{1}{2}\hat{g}_{AC}+\hat{g}_{AA}+
	\frac{1}{2}\hat{g}_{AB}) -
	\hat{\mu}_1(\frac{a_{13}}{2}\hat{g}_{AC}+a_{11}\hat{g}_{AA}+
	\frac{a_{12}}{2}\hat{g}_{AB})
	\\ & \hspace{4.5cm}
	- \mu_2(\frac{a_{23}}{2}\hat{g}_{AC}+a_{21}\hat{g}_{AA}+
	\frac{a_{22}}{2}\hat{g}_{AB}),
	\\ \nonumber &
	\frac{\partial\log Q}{\partial
	\hat{\omega}_B(-\mathbf{k})}
	=\frac{1}{Q} \frac{\partial Q}{\partial
	\hat{\omega}_B(-\mathbf{k})} \thickapprox
	\hat{\mu}_+(\frac{1}{2}\hat{g}_{BC}+\hat{g}_{BB}+
	\frac{1}{2}\hat{g}_{AB}) -
	\hat{\mu}_1(\frac{a_{13}}{2}\hat{g}_{BC}+a_{12}\hat{g}_{BB}+
	\frac{a_{11}}{2}\hat{g}_{AB}) \\ &\hspace{4.5cm}-
	\mu_2(\frac{a_{23}}{2}\hat{g}_{BC}+a_{22}\hat{g}_{BB}+
	\frac{a_{21}}{2}\hat{g}_{AB}),
	\\ \nonumber &
	\frac{\partial\log Q}{\partial
	\hat{\omega}_C(-\mathbf{k})}
	=\frac{1}{Q} \frac{\partial Q}{\partial
	\hat{\omega}_C(-\mathbf{k})} \thickapprox
	\hat{\mu}_+(\frac{1}{2}\hat{g}_{AC}+\hat{g}_{CC}+
	\frac{1}{2}\hat{g}_{BC}) -
	\hat{\mu}_1(\frac{a_{11}}{2}\hat{g}_{AC}+a_{13}\hat{g}_{CC}+
	\frac{a_{12}}{2}\hat{g}_{BC}) \\ &
	\hspace{4.5cm}-\hat{\mu}_2(\frac{a_{21}}{2}\hat{g}_{AC}+a_{23}\hat{g}_{CC}+
	\frac{a_{22}}{2}\hat{g}_{BC}).
\end{align}
Summing them up, 
\begin{align}
	\nonumber
	\frac{\partial\log Q}{\partial
	\hat{\mu}_+(-\mathbf{k})} &=
	\frac{\partial\log Q}{\partial \hat{\omega}_A(-\mathbf{k})} +
	\frac{\partial\log Q}{\partial\hat{\omega}_B(-\mathbf{k})} +
	\frac{\partial\log Q}{\partial\hat{\omega}_C(-\mathbf{k})}
	\\
	& \thickapprox
	\hat{\mu}_+(\mathbf{k})\left(\hat{g}_{AA}+ \hat{g}_{BB}+\hat{g}_{CC}+
	\hat{g}_{AB}+\hat{g}_{AC}+\hat{g}_{BC}\right)
	+ o(\hat\mu_1) + o(\hat\mu_2).
	\label{}
\end{align}
We also note
\begin{align}
\hat{g}_{AA}+
\hat{g}_{BB}+\hat{g}_{CC}+
\hat{g}_{AB}+\hat{g}_{AC}+\hat{g}_{BC} = \hat{g}(\mathbf{k}^2,1)
= \hat{G}(\mathbf{k}),
\end{align}
and the energy functional of $ABC$ triblock copolymers in
Fourier-space is
\begin{align}
	H = -\hat{\mu}_{+}(\mathbf{0}) +
	\frac{1}{4N\beta_1}\sum\limits_{\mathbf{k}}\hat{\mu}_{1}(\mathbf{k})\,\hat{\mu}_{1}(-\mathbf{k})
	+\frac{1}{4N\beta_2}\sum\limits_{\mathbf{k}}\hat{\mu}_{2}(\mathbf{k})\,\hat{\mu}_{2}(-\mathbf{k})-\log\hat{q}(\mathbf{0},1).
	\label{eqn:triblock:hamiltonian:fourier}
\end{align}
Thus 
\begin{align}
	\widehat{\Bigg(\frac{\delta^2
	H}{\delta\mu_+^2}\Bigg)}(\bm k) = -
\frac{2}{\mathbf{k}^4}(e^{-\mathbf{k}^2}+\mathbf{k}^2-1
) \stackrel{\Delta}{=} -\hat{G}(\bm k),
\end{align}
where the caret denotes Fourier transform. 
In fact, the above result is hold for general linear
multi-block copolymers by the same derivative process.

Following the above derivative process, 
for general nonlinear polymer chain, the approximation
second-order variation of $H$ with respect to $\mu_+$ in
Fourier space is 
\begin{align}
	\widehat{\Bigg(\frac{\delta^2
	H}{\delta\mu_+^2}\Bigg)}(\bm k) \approx
	- \sum_{\alpha} \hat{g}_{\alpha\alpha}(\mathbf{k}^2)
	\stackrel{\Delta}{=} -\hat{G}(\bm k),
	\label{eqn:method:sis:debyestar}
\end{align}
where
$\hat{g}_{\alpha\alpha}(\mathbf{k})=2(e^{-\mathbf{k}f_\alpha}+\mathbf{k}f_\alpha-1)/{\mathbf{k}^2}$
is the familiar Debye
function\,\cite{fredrickson2006equilibrium}, $\alpha$ denotes the
$\alpha$-th species.

\end{appendix}

\end{document}